%% file: Particle_Filtering_and_Gaussian_Mixtures_Rojahn-et-al_v1.tex
\documentclass{article}
\include{mysty}
\usepackage{arxivnew}

\usepackage[colorinlistoftodos]{todonotes}
\usepackage{longtable}
\usepackage{setspace}

\usepackage[utf8]{inputenc} % allow utf-8 input
\usepackage[T1]{fontenc}    % use 8-bit T1 fonts
\usepackage{url}            % simple URL typesetting
\usepackage{booktabs}       % professional-quality tables
\usepackage{amsfonts}       % blackboard math symbols
\usepackage{nicefrac}       % compact symbols for 1/2, etc.
\usepackage{microtype}      % microtypography
\usepackage{lipsum}         % Can be removed after putting your text content
\usepackage{graphicx}
\usepackage{natbib}
\usepackage{doi}

\newcommand{\myred}{\color[rgb]{0,0,0}} 
\newcommand{\myblack}{\color[rgb]{0,0,0}}        % changes for revision 1

\usepackage{hyperref}       % hyperlinks
\usepackage{cleveref}       % smart cross-referencing

\title{Particle Filtering and Gaussian Mixtures - \\ On a Localized Mixture Coefficients Particle Filter (LMCPF) for global NWP}

\renewcommand{\shorttitle}{Particle Filtering and Gaussian Mixtures}

\pdfoutput=1
\hypersetup{
pdftitle={Particle Filtering and Gaussian Mixtures - On a Localized Mixture Coefficients Particle Filter (LMCPF) for global NWP},
pdfauthor={Anne~Rojahn, Nora~Schenk, Peter Jan~van Leeuwen, Roland~Potthast},
pdfkeywords={data assimilation, high dimensional, particle filter, Non-Gaussian, numerical weather prediction},
}

\begin{document}
\maketitle
%{\Large
%\textbf\newline{Particle Filtering and Gaussian Mixtures - \\ On a 
%Localized Mixture Coefficients Particle Filter (LMCPF) for global NWP}
%}
%\newline
%% authors go here:
%\\
%Anne Walter\textsuperscript{1,*},
%Nora Schenk\textsuperscript{1,2},
%Peter Jan van Leeuwen\textsuperscript{3},
%Roland Potthast\textsuperscript{1}
%\\
%%\bigskip
%\\
%\bf{1} Deutscher Wetterdienst
%\\
%\bf{2} Affiliation B
%\\
%\bf{3} Affiliation B
%\\
%%\bigskip
%\\
%* Anne.Walter@dwd.de

\begin{abstract}
In a global numerical weather prediction (NWP) modeling framework we study the implementation of {\em Gaussian uncertainty} of individual particles into the assimilation step of a localized adaptive particle filter (LAPF). We obtain a local representation of the prior distribution as a mixture of basis functions. In the assimilation step, the filter calculates the individual weight coefficients and new particle locations. It can be viewed as a combination of the LAPF and a localized version of a Gaussian mixture filter, i.e., a {\em Localized Mixture Coefficients Particle Filter} (LMCPF). 

Here, we investigate the feasibility of the {\myred LMCPF} within a global operational framework and evaluate the relationship between prior and posterior  distributions and observations. Our simulations are carried out in a standard pre-operational experimental set-up with the full global observing system, 52 km global resolution and $10^6$ {\myblack model variables}. Statistics of particle movement in the assimilation step are calculated. The mixture approach is able to deal with the discrepancy between prior distributions and observation location in a real-world framework and to pull the particles towards the observations in a much better way than the pure {\myred LAPF}. This shows that using Gaussian uncertainty can be an important tool to improve the analysis and forecast quality in a particle filter framework.
\end{abstract}

% keywords can be removed

\section{Introduction}
Let us consider a state space $\R^n$ of dimension $n\in \N$ , an observation space $\R^m$ of dimension $m \in \N$ and a sequence of observations $y_k \in \R^m$ at points in time $t_k$ for time index $k=1,2,3,\dots$. 
Based on a {\em prior} distribution $p^{(b)}_{k}(x)$, $x \in \R^n$, at time $t_k$, the task of {\em Bayesian data assimilation} is to calculate a {\em posterior} probability distribution $p^{(a)}_{k}(x)$, $x\in \R^n$, at time $t_k$. States and observations are linked by the equation
\beq
{\myred y_k = H(x_k^{true})+\epsilon_k}
\eeq 
{\myred with the true state vector $x_k^{true} \in \R^n$ at time $t_k$, some observation error $\epsilon_k$} and the {\em observation operator} $H: \R^n \rightarrow \R^m$. Usually, the prior $p^{(b)}_{k}$ is estimated from earlier analysis steps, from which the distribution is propagated through time to some recent analysis time $t_{k}$ based on some numerical model $M$. 
%Both numerical approximation error as well as model error leads to errors in the estimate of $p^{(b)}_{k}$.

The approximation of a general prior distribution by an {\em ensemble of states}, also known as a set of {\em particles}, has a long tradition in mathematical stochastics, see for example \cite{Crisan1}. It is also well-known, that sampling as usually carried out by Markov Chain Monte Carlo (MCMC) methods \citep{Anderson99,Crisan1,Crisan2} works well in low dimensions, but when we sample in a high-dimensional space (where high usually refers to dimensions above n=5), the methods basically collapse, since the number of necessary samples  to find some probability different from zero grows exponentially with the dimension \citep{Leeuwen10,Snyder08, Snyder15,Bickel08}. Alternative methods based on particular approximations of the prior and posterior have been developed, with the {\em Ensemble Kalman Filter} (EnKF) \citep{Evensen,even00,Evensen2} and the {\em Local Ensemble Transform Kalman Filter} (LETKF) by \cite{Hunt} as important and widely used methods for high-dimensional filtering. These methods, however, rely on the approximation of the prior by a Gaussian, which is a strong limitation when applied to highly non-linear dynamical systems as either global or high-resolution {\em Numerical Weather Prediction} (NWP). 

Different routes to carry out non-Gaussian assimilation have been taken by the filtering community for example with Gaussian mixtures \citep{Anderson99}, locally applied Gaussian mixtures \citep{Bengtsson03} or by the development of particular filters such as the GIGG filter of \cite{bishop16}. For an overview of different ensemble-based data assimilation methods, we refer to \cite{vetra} and \cite{Leeuwen19}. An alternative route has been chosen by the 4D-VAR community with an ensemble of 4D-VARS based on perturbed observations, compare \cite{ecmwf}. 

Over the past years {\em particle filters} have become mature enough to be used for very-high-dimensional non-Gaussian filtering, compare \cite{Leeuwen09,Leeuwen,Farchi18} and \cite{Leeuwen19} for recent reviews. {\em Localization} for particle filters is used by \cite{Reich,poterjoy,Penny16} and \cite{Potthast19}. {\myblack Instead of the localization \cite{kawabata2020} have used an adaptive observation error estimator to avoid the filter collapse in a regional mesoscale model.} Particle filters have been successfully used for full-scale NWP systems. In particular, in \cite{poterjoy17} a localized particle filter has been studied for a regional numerical weather prediction model over the US. The team \cite{FK13} developed a {\em hybrid Ensemble Kalman Particle Filter} which \cite{Robert2017} has tested for the regional COSMO NWP model. The {\em Localized Adaptive Particle Filter} (LAPF) described in \cite{Potthast19} has been tested for the global ICON NWP model. The LAPF \citep{Potthast19} has shown to provide reasonable assimilation results for an global atmospheric data assimilation for the ICON model in quasi-operational setup. It has been successfully run for a month of assimilations with $10^6$ degrees of freedom (52 km global resolution) and shows a stable behaviour synchronizing the system with reality. 

Here, our starting point is the investigation of the behaviour of the LAPF with respect to errors in the prior distribution $p^{(b)}_{k}$. By studying the statistics of the observations vector mapped into ensemble space, we will show that in many cases the model forecasts show significant distance to the observations, and the particle filter based on a limited number of delta distributions does not pull the particles close enough to the observations when the move of particles is only achieved through adaptive resampling. 

To allow individual particles to move towards the observations, we further develop the LAPF by bringing ideas from Gaussian mixtures into its framework. We reach this goal by including model and forecast uncertainty for each particle, as for example suggested by the {\em Low-Rank Kernel Particle Kalman Filter} (LRKPKF) of \cite{Hoteit08}, compare also \cite{liu2016} and \cite{liu2016b}. The basic idea is to consider each particle to be a Gaussian where its width is representing its uncertainty. This means we study a prior distribution given by a Gaussian (or more general radial basis function RBF) mixture. Then, the prior has the form
\beq
p^{(b)}(x) := c  \sum_{\ell=1}^{L} c_{\ell} e^{-\frac{1}{2} (x-x^{(b,\ell)})^{T}
\bG^{-1}(x-x^{(b,\ell)})} , \;\; x \in \R^n, 
\label{Gaussian mixture 0}
\eeq
with constants {\myred $c_{\ell}= 1/\sqrt{(2 \pi)^n \det(\bG)}$} for the individual Gaussian basis functions with mean $x^{(b,\ell)}$ and covariance $\bG$ and a normalization constant $c${\myblack, which in this case is given by $c=1/L$, but is used as a generic constant which is different in different equations.} For this approximation, and when the observation operator $H$ is linear, we can explicitly calculate the posterior distribution as a corresponding Gaussian mixture, i.e.,\
\beq
p^{(a)}(x) := c \sum_{\ell=1}^{L} c_{\ell} w_{\ell} 
e^{-\frac{1}{2} (x-\tx^{(a,\ell)})^{T}
\tilde{\bG}^{-1}(x-\tx^{(a,\ell)})}, \;\; x \in \R^n, 
\eeq
with some matrix $\tilde{\bG}$ (calculated e.g.\ in Chapter 5.4 of \cite{NakaPot}), constants $c_{\ell} w_{\ell}$ {\myred (not normalized at this point, since different Gaussians get different weight in this first Bayesian step)} and temporary analysis states $\tx^{(a,\ell)}$, $\ell=1,...,L$, with the components
\beq
{\myblack q^{(a,\ell)}(x)} := c c_{\ell} w_{\ell} 
e^{-\frac{1}{2} (x-\tx^{(a,\ell)})^{T}
\tilde{\bG}^{-1}(x-\tx^{(a,\ell)})}, \;\; x \in \R^n. 
\eeq 
{\myred The constant $c$ will normalize the integral of $p^{(a)}$ to one.}
If there are no further constraints to the variables, the $\ell$-th posterior particle can be directly drawn {\myred with relative probability $c_{\ell} w_{\ell}$} from the distribution {\myblack component $q^{(a,\ell)}(x)$} leading to an analysis ensemble member $x^{(a,\ell)}$. This drawing process is carried out based on {\em localization}, {\em adaptivity} and the {\em transformation into ensemble space} as developed for the LAPF \citep{Potthast19}; details will be described in Sections \ref{lapf} and \ref{filtering step}. {\myred As for other particle filters, the posterior particles will be calculated by an ensemble transform matrix, with details worked out in Section \ref{filtering step}. For each posterior ensemble member, based on the prior Gaussian mixture, this matrix defines {\em coefficients} describing the weights of of each particle.} The name {\em Local Mixture Coefficients Particle Filter (LMCPF)} has been used {\myred to distinguish from other localized particle filter methods. For example, \cite{Reich} present Localized Particle Filter (LPF) versions, which include sophisticated optimal transport properties. A further LPF method is introduced by \cite{Penny16} and the LAPF (already implemented at the German Weather Service in 2014\footnote{Shown by German Climate Computing Center DKRZ Git Records}) is presented by \cite{Potthast19}.} We note that the choice for $\bG$ of formula (\ref{Gaussian mixture 0}) as a scaled version of Hunt the ensemble correlation matrix, i.e.,\ {\myred$\bG = \kappa \bB$, with $\bB = \frac{1}{(L-1)} \bX\bX^T$}, resembles the choices made for the LETKF \citep{Hunt} and leads to very efficient code. 

We will investigate the usefulness of the Gaussian uncertainty within the particle filter in very high-dimensional systems, leading to moves or shifts of the particles towards the observations. Statistics of these shifts will be shown, demonstrating that for this global atmospheric NWP system the uncertainty plays an important role. Further, our numerical results show that the LMCPF is a particle filter with a quality comparable to the LETKF for state-of-the-art real-world operational global atmospheric NWP forecasting systems. This will be demonstrated by numerical experiments based on an implementation of the particle filter in the operational data assimilation software suite DACE\footnote{Data Assimilation Coding Environment} of Deutscher Wetterdienst (DWD). 

The {\em Localized Mixture Coefficients Particle Filter} is introduced in Section \ref{lmcpf}, where we first summarize the ingredients we build on in Section \ref{lapf}. Then, an elementary Gaussian filtering step in ensemble space is described in Section \ref{filtering step}. Finally, the full LMCPF method is presented in Section \ref{lmcpf full}. {\myred We describe the high-dimensional experimental environment for our development and evaluation framework for numerical tests in Section \ref{environment}.} The numerical results for the global weather forecasting model ICON are shown in Section \ref{numerics}. We study the statistics of the relationship of observations and the ensemble as well as the corresponding statistics of the shift vectors of the Gaussian particles of the LMCPF. We show the large improvements with respect to standard NWP scores which the LMCPF can achieve compared to the LAPF. Additionally, we present case studies comparing the LMCPF forecast scores to the operational LETKF.

%===============================================================================
%
%
%===============================================================================
\section{Localized Mixture Coefficients Particle Filter (LMCPF)}
\label{lmcpf}

The basic idea of a Bayesian assimilation step is to calculate a posterior
distribution $p^{(a)}(x)$ for a state $x \in \R^n$
based on a prior distribution $p^{(b)}(x)$ for $x\in \R^n$, some measurement
$y \in \R^m$ and a distribution of the measurement error $p(y|x)$ of $y$ given the 
state $x$. The famous Bayes formula calculates 
\begin{equation}
p^{(a)}(x) = c p^{(b)}(x) \cdot p(y|x), \;\; x \in \R^n, 
\label{Bayes}
\end{equation}
with normalization constant $c$ such that $\int_{\R^n} p^{(a)}(x) \; dx = 1$. 

Our setup for data assimilation is to employ an ensemble $\{x^{(b,\ell)} \in \R^n, \ell=1,...,L\}$ of  states, which are used to estimate or approximate $p^{(b)}(x)$. The basic analysis step of data assimilation is to construct an analysis ensemble $\{ x^{(a,\ell)} \in \R^n, \ell=1,...,L\}$ of {\em analysis states}, which approximate $p^{(a)}(x)$ in a way consistent with the approximation of $p^{(b)}(x)$ by $x^{(b,\ell)}$, $\ell=1,...,L$. The above idea is common to both the Ensemble Kalman Filter (EnKF) and to particle filters. We employ the notation
\beq
\bX^{(b)} := \Big( x^{(b,1)}-\ox, ..., x^{(b,L)}-\ox \Big) \in \R^{n\times L}
\label{X}
\eeq
for the matrix of ensemble differences to the ensemble mean $\ox$ {\myblack  defined by
\beq
\ox := \frac{1}{L} \sum_{\ell=1}^{L} x^{(b,\ell)} \in \R^n. 
\eeq 
For the ensemble differences in observation space we employ
\beq
\bY^{(b)} := \Big( y^{(b,1)}-\oy, ..., y^{(b,L)}-\oy \Big) \in \R^{m\times L} 
\label{Y}
\eeq
with the mean $\oy$ defined by
\beq
\oy := \frac{1}{L} \sum_{\ell=1}^{L} y^{(b,\ell)} \in \R^m
\label{ybar}
\eeq 
and
\beq
y^{(b,\ell)} := H(x^{(b,\ell)}).
\eeq
From now on we will use $\bX$ for $\bX^{(b)}$ and $\bY$ for $\bY^{(b)}$ for brevity.} In the case of a linear observation operator we have $\oy = \bH \ox$ and $\bY = \bH\bX$. Usually, for Ensemble Kalman Filters, the approximation {\myblack of the covariance matrix} is chosen to be based on the estimator
\begin{equation}
\bB := \frac{1}{L-1} \sum_{\ell=1}^{L} (x^{(\ell)}-\ox) \cdot (x^{(\ell)}-\ox)^{T} \in \R^{n \times n}. 
\end{equation}
{\myblack The estimator} {\myred $\bB$} can also be written as $\bB = \frac{1}{L-1} \bX \bX^{T}$.
Usually, in this case the prior is approximated by 
\beq
p^{(b)}(x) = c_{B} e^{-\frac{1}{2} (x-\ox)^{T} \bB^{-1} (x-\ox)}
\eeq
{\myred with $\bB^{-1}$ well defined\footnote{\myred{The standard arguments, see Lemma 3.2.1 of \cite{NakaPot}, show injectivity of $XX^T$ on $R(X)$: $XX^T X\beta = 0$ with $\beta \in\R^L$ yields $X^T X \beta \in N(X) \cap R(X^T) = R(X^T)^{\perp} \cap R(X^T)$, thus $X^T X \beta = 0$. The same argument for $X\beta \in N(X^T)$ yields $X\beta=0$, thus $X X^T$ is injective on $R(X)$. For surjectivity we consider $v \in R(X)$, i.e. $v = X w$ with $w \in \R^L = N(X) \oplus N(X)^{\perp} = N(X) \oplus R(X^T)$, such that $w = w_1 + w_2$ with $w_1 \in N(X)$ and $w_2 = X^T \beta$ with some $\beta \in \R^n = R(X) + R(X)^{\perp}$. Repeating the last argument leads to a $\beta_1 \in R(X)$ with $w=X^T \beta_1$ and thus surjectivity. Invertibility of $B$ is thus shown.}  
} for all $x = \ox + \bX \beta$ with some vector $\beta \in \R^{L}$. The normalization constant $c_B$ can be calculated based on a matrix $\Phi$ which consists of an orthonormal basis of $N(X)^{\perp} \subset \R^L$ of dimension $\tL < L$ by 
\beq
c_{B} := \left( \int_{\R^{\tL}} e^{-\frac{1}{2} (X\Phi\alpha)^{T} \bB^{-1} (X\Phi\alpha)} \sqrt{\det(\Phi^{T}X^{T}X\Phi)} \; d\alpha \right)^{-1}, 
\eeq 
where $\det(\Phi^{T}X^{T}X\Phi)$ is the Gramian of the injective mapping $X\Phi: \R^{\tL} \rightarrow \R^n$, i.e.\ the determinant of the Gram matrix $\Phi^{T}X^TX\Phi$.} The approximation of the {\em classical particle filter} is
\beq
p^{(b)}(x) = c \sum_{\ell=1}^{L} \delta(x-x^{(b,\ell)}), \;\; x \in \R^n,
\eeq
with the {\em delta distribution} $\delta(\cdot)$ and a normalization constant $c=1/L$. A well-known
idea is to employ {\em Gaussian mixtures} (c.f.\ \cite{Hoteit08,liu2016,liu2016b}), i.e.,\ use the approximation
\beq
p^{(b)}(x) = c \sum_{\ell=1}^{L} c_{\ell} e^{-\frac{1}{2} (x-x^{(b,\ell)})^{T} {\myblack \bG_{\ell}}^{-1} (x-x^{(b,\ell)})},
\label{Gaussian mixture}
\eeq
where {\myblack $\bG_{\ell} \in \R^{n\times n}$} is some symmetric and positive definite matrix which describes the 
{\em uncertainty} of the individual particle, {\myred $c_{\ell}=  1/\sqrt{(2 \pi)^n \det(\bG_{\ell})}$} is a normalization constant for each
of the Gaussians under consideration and $c$ is an overall normalization constant. 
\begin{itemize}
\item
The matrix {\myblack $\bG_{\ell}$} is the covariance of each Gaussian and can be seen as a measure for the short-range {\em forecast error} consisting of model error and some of the uncertainty in the initial conditions beyond the distribution of the ensemble of particles itself. We will discuss the important role of {\myblack $\bG_{\ell}$} in several places later, when we describe the LMCPF and its numerical realization. {\myred In particular, we will investigate the situation where $G$ is a multiple of the covariance matrix $B$ defined above.}
\item
The Gaussian mixture filter can be seen as a generalization of
the classical particle filter, where instead of a delta distribution a Gaussian around each prior particle is employed to calculate the posterior distribution and draw from it. Here, we will employ {\em localization} and {\em adaptivity} as developed for the LAPF in combination with the mixture concept within the LMCPF. 
\end{itemize}

%===============================================================================
%
%
%===============================================================================
\subsection{The Localized Adaptive Particle Filtering Ingredients and Preparations}
\label{lapf}

The goal of this section is to collect, prepare and summarize all components employed for the {\em localized mixture coefficients particle filter}. For the following derivation we assume linearity of $\bH$, we will discuss the form of the equations in the case of non-linear $\bH$ later. Then, we have $\bY^{T} = \bX^{T} \bH^{T}$ and with $\gamma = \frac{1}{L-1}$ the standard estimator for the covariance matrix is given by $\bB = \gamma \bX \bX^{T}$. We will later use $\bB$ as measure of uncertainty of individual particles, then using the scaling 
\begin{equation}
\label{kappa def}
\gamma = \frac{\kappa}{(L-1)}
\end{equation} 
with a parameter $\kappa>0$ scaling the standard covariance matrix. Following standard arguments as in \cite{Hunt,NakaPot} or \cite{Potthast19}, this leads to the
{\em Kalman gain}
\bea
\bK & = & \bB \bH^{T}(\bR+\bH\bB\bH^{T})^{-1} \nonumber \\
& = & \gamma \bX \bX^{T} \bH^{T}( \bR + \gamma \bH \bX \bX^{T} \bH^{T})^{-1} \nonumber \\
& = & \gamma \bX \bY^{T}(\bR + \gamma \bY \bY^{T})^{-1} 
\label{deriv101}
\eea
with invertible observation error covariance matrix $\bR \in \R^{m\times m}$. We note that we have
\beq
(\bI + \gamma \bY^{T} \bR^{-1} \bY) \bY^{T} = \bY^{T} \bR^{-1} (\bR + \gamma \bY \bY^{T})
\label{deriv102a}
\eeq
by elementary calculations. We also note that $\bI + \gamma \bY^{T} \bR^{-1} \bY$
is invertible on $\R^L$ and $\bR + \gamma \bY \bY^{T}$ is invertible on 
$\R^m$ by assumption on the invertibility of $\bR$. Then, multiplying
(\ref{deriv102a}) by $(\bI + \gamma \bY^{T} \bR^{-1} \bY)^{-1}$ from the left
and by $(\bR + \gamma \bY \bY^{T})^{-1}$ from the right we obtain
\beq
\bY^{T}(\bR + \gamma \bY \bY^{T})^{-1} = (\bI + \gamma \bY^{T} \bR^{-1} \bY)^{-1}
\bY^{T} \bR^{-1}.
\label{deriv103} 
\eeq
Now, (\ref{deriv103}) can be used to transform (\ref{deriv101}) into
\bea
\label{deriv101b}
\bK = \gamma \bX (\bI + \gamma \bY^{T}\bR^{-1} \bY)^{-1} \bY^{T}\bR^{-1}. 
\eea
This can be used to calculate the covariance update step of the
Kalman filter in ensemble space as follows. We derive
\bea
\bB^{(a)} & = & (\bI - \bK\bH) \bB^{(b)} \nonumber \\
& = & \Big( \bI - \gamma \bX (\bI + \gamma \bY^{T}\bR^{-1} \bY)^{-1} \bY^{T}\bR^{-1} \bH \Big) 
\gamma \bX \bX^{T} \nonumber \\
& = & \bX \Big( \bI - \gamma (\bI + \gamma \bY^{T}\bR^{-1} \bY)^{-1} \bY^{T}\bR^{-1} \bY \Big) 
\gamma \bX^{T} \nonumber \\
& = & 
\bX \Big( (\bI + \gamma \bY^{T}\bR^{-1} \bY)^{-1} \Big[ \bI + \gamma \bY^{T} \bR^{-1} \bY
- \gamma \bY^{T} \bR^{-1} \bY \Big]\Big) \gamma \bX^{T} \nonumber \\
& = &  \bX (\bI + \gamma \bY^{T}\bR^{-1} \bY)^{-1}  \gamma \bX^{T}
\nonumber \\
& = & \gamma \bX (\bI + \gamma \bY^{T} \bR^{-1} \bY)^{-1} \bX^{T}. 
\label{deriv102}
\eea
The analysis ensemble $\bX^{(a)}$ which generates the correct posterior
covariance by $\bB^{(a)} = \gamma \bX^{(a)} (\bX^{(a)})^T$ is given by 
\beq
\bX^{(a)} := \bX^{(b)} \Big(\bI + \gamma \bY^{T}\bR^{-1}\bY \Big)^{-\frac{1}{2}} \in \R^{{\myblack n}\times L}, 
\eeq
where the matrix $\bI + \gamma \bY^{T}\bR^{-1}\bY \in \R^{L \times L}$
lives in ensemble space, it is symmetric and invertible by construction,
for all $\gamma > 0$.

The localized ensemble transform Kalman filter (LETKF) following 
\cite{Hunt} based on the square root filter for calculating the
analysis ensemble can be written as
\bea
\ox^{(a)}  :=  
\ox^{(b)} + \gamma \bX^{(b)} w 
= \ox^{(b)} + \bK (y-\oy)
\eea
with
\beq
w := (\bI + \gamma \bY^{T}\bR^{-1} \bY)^{-1} \bY^{T}\bR^{-1} (y-\oy) \in \R^{L}
\eeq 
and 
\beq
\label{ens trans}
\bX^{(a)} := \bX^{(b)} \bW
\eeq
with 
\beq
\label{W def}
\bW := (\bI + \gamma \bY^{T}\bR^{-1} \bY)^{-\frac{1}{2}} \in \R^{L\times L}.
\eeq
{\myblack The above equations are carried out at each analysis grid point where the
matrix $\bR$ is localized by multiplication of each entry with a localization function depending on the distance of the variable to the analysis grid point \cite{Hunt}.}
Using
\beq
\bX^{(a,full)} := \Big( x^{(a,1)}, ..., x^{(a,L)} \Big) = (\ox^{(a)} + x^{(a)}) \in \R^{n\times {\myblack L}}
\label{Xa full}
\eeq
the full update of the LETKF ensemble can be written as
\beq
\bX^{(a,full)} = \ox^{(b)} + \gamma \bX^{(b)} w + \bX^{(b)} \bW, 
\label{LETKF final} 
\eeq 
where we define the sum of a vector (here $\ox^{(b)}$ or $\gamma \bX^{(b)} w$)
plus a matrix (here $\bX^{(b)}\bW)$ by adding the vector to
each column of the matrix. 

{\myblack For non-linear observation operator $H$ as in (18) of \cite{Hunt} the operator $\bK$ is defined by the last line of (\ref{deriv101}), see also (\ref{deriv101b}) and the ensemble transform by (\ref{ens trans}) with $\bW$ by (\ref{W def}). This basically corresponds to an approximate linearization of $H$ in observation space based on the differences $y^{(b,\ell)} - \oy$.} 

%===============================================================================
%
%
%===============================================================================
\subsection{An Elementary Gaussian Filtering Step in Ensemble Space}
\label{filtering step}

Let us consider a Bayesian assimilation step (\ref{Bayes}) based on the approximation
of the prior $p^{(b)}(x)$ as a Gaussian mixture (\ref{Gaussian mixture}). 
We first describe the steps in general, then derive the ensemble space version of
the equations. To each particle, we attribute a distribution with covariance $\bG$, i.e.,\
we define
\beq
p^{(b,\ell)}(x) := \frac{1}{\sqrt{(2\pi)^n \det(\bG)}}
e^{-\frac{1}{2} (x - x^{(b,\ell)})^{T} \bG^{-1} (x - x^{(b,\ell)})}, \;\;
x \in \R^n, 
\label{pp}
\eeq
which is normalized according to equation (4.5.28) of \cite{NakaPot}. Then, 
the full prior is a {\em Gaussian mixture}
\beq
\label{prior 1}
p^{(b)}(x) = c \sum_{\ell=1}^{L} c_{\ell}
e^{-\frac{1}{2} (x - x^{(b,\ell)})^{T} \bG^{-1} (x - x^{(b,\ell)})}, \;\;
x \in \R^n, 
\eeq
with $c_{\ell} := 1/\sqrt{(2\pi)^n \det(\bG)}$ (i.e.,\ we choose the variance uniform for all $\ell$) and
with some normalization constant $c=\frac{1}{L}$ in this case. 
Bayes formula leads to the posterior distribution
\beq
p^{(a)}(x) = c \sum_{\ell=1}^{L} c_{\ell} \Big( 
e^{-\frac{1}{2} (x-x^{(b,\ell)})^{T} \bG^{-1} (x-x^{(b,\ell)})}
e^{-\frac{1}{2} (y-H(x))^{T} \bR^{-1} (y-H(x))} \Big), 
\label{posterior 1}
\eeq
$x \in \R^n$, 
with a normalization constant $c$, here different from the normalization constant in (\ref{prior 1}). We note that the terms in round brackets constitute individual Gaussian assimilation steps. The posterior of each of these terms can be explicitly calculated the same way as for the Ensemble Kalman Filter. Following
\cite{NakaPot}, Section 5.4, we define
\beq
x^{(a,\ell)} := x^{(b,\ell)} + \bG \bH^{T} ( \bR + \bH \bG \bH^{T})^{-1}(y-H(x^{(b,\ell)})), \;\; \ell=1,...,L,
\label{shift0}
\eeq
and
\beq
\bK = \bG \bH^{T}(\bR + \bH \bG \bH^{T})^{-1}, \;\; \bG^{(a)} := (I-\bK \bH) \bG. 
\label{post0}
\eeq
Then, we know that
\bea
q^{(a,\ell)}(x) & := & c_{\ell}
e^{-\frac{1}{2} (x-x^{(b,\ell)})^{T} \bG^{-1} (x-x^{(b,\ell)})}
e^{-\frac{1}{2} (y-H(x))^{T} \bR^{-1} (y-H(x))} 
\nonumber \\
& = &  w_{\ell} e^{-\frac{1}{2}(x-x^{(a,\ell)})^{T} [\bG^{(a)}]^{-1} (x-x^{(a,\ell)})}, \;\; x \in \R^n, 
\eea
with constants $w_{\ell}$ given by
\bea
w_{\ell} & = & \int_{\R^n} c_{\ell} e^{-\frac{1}{2} (x-x^{(b,\ell)})^{T} \bG^{-1} (x-x^{(b,\ell)})}
e^{-\frac{1}{2} (y-H(x))^{T} \bR^{-1} (y-H(x))}  \; dx \nonumber \\
& & \cdot \frac{1}{\sqrt{ (2\pi)^n \det(\bG^{(a)}) }}. 
\label{wl full}
\eea
Since both $c_{\ell}$ and $\sqrt{ (2\pi)^n \det(\bG^{(a)}) }$ do not depend on 
$\ell$, the constants are irrelevant for the resampling step and will be removed by the 
normalization step.
Note that the constants $w_{\ell}$, $\ell=1,...,L$, are extremely
important, since they contain the relative weights of the individual posterior particles
with respect to each other. They should not be ignored!  
Here, we first describe the full posterior distribution, which is now given by
\beq
p^{(a)}(x) = c \sum_{\ell=1}^{L} w_{\ell} 
e^{-\frac{1}{2}(x-x^{(a,\ell)})^{T} [\bG^{(a)}]^{-1} (x-x^{(a,\ell)})}, \;\; x\in \R^n.
\eeq
In the case of the classical particle filter, the Gaussians
$c_{\ell} e^{-\frac{1}{2} (x-x^{(b,\ell)})^{T} \bG^{-1} (x-x^{(b,\ell)})}$ become $\delta$-distributions
$c_{\ell} \delta(x-x^{(b,\ell)})$ with weights $c_{\ell} = 1$. In this case, 
the individual posterior weights $w_{\ell}$ are given by the likelihood of observations
\beq
w_{\ell} := e^{-\frac{1}{2}(y-H(x^{(b,\ell)}))^{T} \bR^{-1} (y-H(x^{(b,\ell)}))}, \;\; \ell=1,...,L. 
\label{wl pf}
\eeq
This choice will also be a reasonable approximation in the case of small variance
$\bG$ of the Gaussians under consideration in comparison with the distance 
$y-H(x^{(b,\ell)})$. 
In the general Gaussian case, the weights can be calculated from (\ref{wl full}). {\myblack For our
numerical experiments we use $G$ non-zero with some positive variance, but approximate $w_l$ by (\ref{wl pf}). }
{\myred 

\begin{figure}
\centering
\includegraphics[width=12cm]{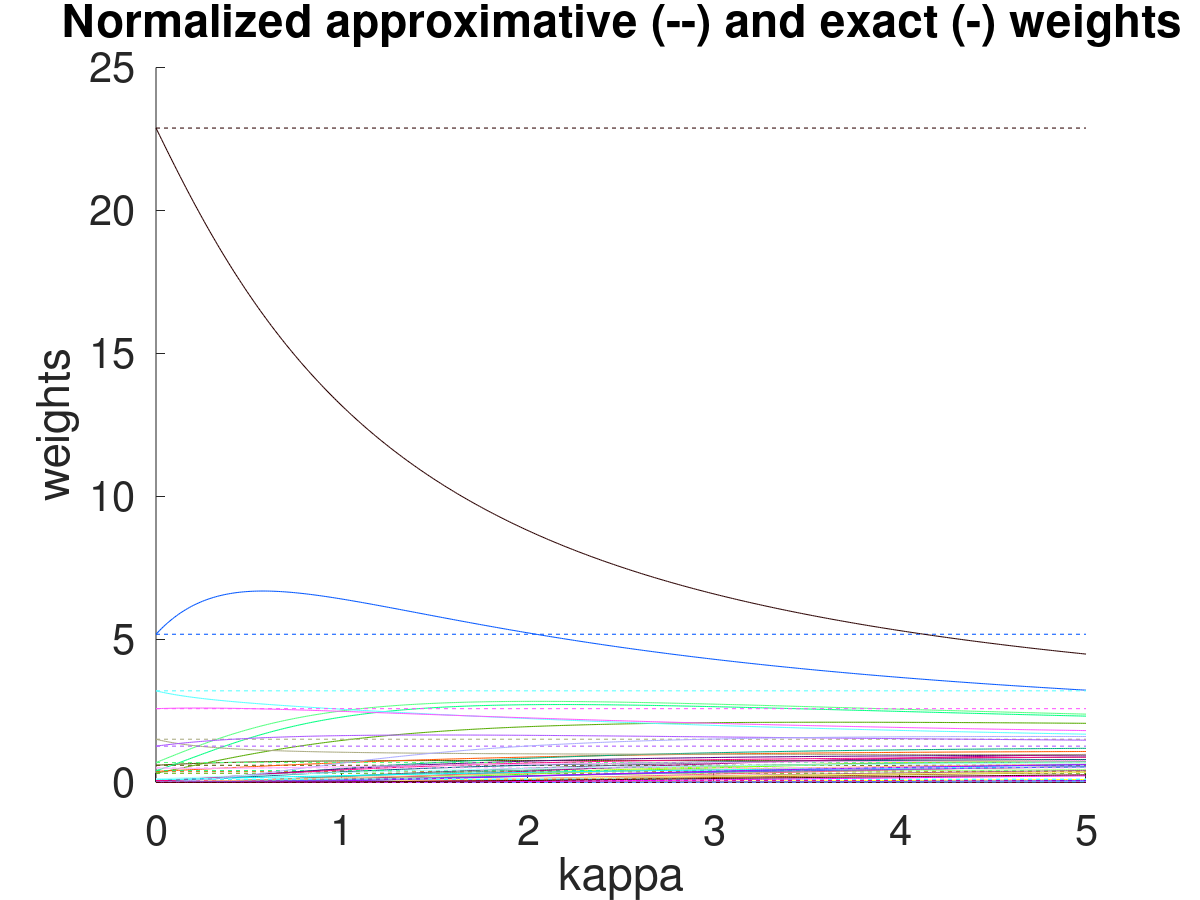}
\caption[We show a comparison between normalized approximative and exact weights.]{
We show a comparison between the normalized approximative weights calculated as in (\ref{wl pf}) versus the normalized exact calculated weights (\ref{wl full}). The solid lines show the normalized exact determined weights and the dashed lines the normalized approximative weights. The colors vary for different ensemble members (L=40). On the x-axis we show the value for $\kappa$ of equation (\ref{kappa def}), on the y-axis the values of the weights.}
\label{CompareWeights}
\end{figure}

In Figure \ref{CompareWeights} we show a comparison of the normalized approximative weights (\ref{wl pf}) as dashed lines and the normalized exact determined weights (\ref{wl full}) as solid lines, for a selected point of the full NWP model described in Sections \ref{environment} and \ref{numerics}. Here, each ensemble member (L=40) is described by a different color. For this plot we varied the parameter $\kappa$, described in equation (\ref{kappa def}), between $0$ and $5$. Figure \ref{CompareWeights} shows how the normalized approximative weights differ from the normalized exact weights. The assumptions used to carry out our experiments may not be fully justified. Clearly, further study and experimental work is necessary.}

\bigskip
Let us now describe the ensemble space transformation of the above equations. The ensemble space as a subset of the state space is spanned by $\bX$ given in (\ref{X}). Our {\em ansatz} for the {\em model error} covariance is $\gamma \bX\bX^T$ with some scaling factor $\gamma$. We note that for the LETKF, $\gamma=\frac{1}{L-1}$. Here, $\gamma >0$ can be any real number. We will provide some estimates for what $\gamma$ can be in a global NWP model setup in our numerical part in  Section \ref{numerics}. In the transformed space this leads to the covariance $\gamma \bI \in \R^{L\times L}$ to be used for the ensemble transform version of (\ref{posterior 1}). Recall the ensemble transformation $x - \ox = \bX \beta$, $x^{(\ell)}-\ox = \bX e_{\ell}$ and $x-x^{(\ell)} = \bX (\beta-e_{\ell})$ for $\ell=1,...,L$ {\myblack , where $e_{\ell}$ is the standard unit vector with one in its
$\ell$-th component and zero otherwise} leading to
\bea
(x-x^{(\ell)})^{T} (\gamma \bX \bX^{T})^{-1} (x - x^{(\ell)}) & = &  
{\myred (\beta-e_{\ell})^{T} \gamma^{-1} \bX^{T} (\bX \bX^{T})^{-1} \bX (\beta-e_{\ell})} \nonumber \\
& = & {\myred (\beta-e_{\ell})^{T} \gamma^{-1} \bI (\beta-e_{\ell}).}
\eea
We note that $\bX^{T} (\bX \bX^{T})^{-1} \bX = \bI$ is true only on the subspace $N(\bX)^{\perp}$, but we can employ the arguments used to justify equation (15) of \cite{Hunt} to use the covariance $\gamma^{-1} \bI$ in ensemble space for the prior term. 
For the observation error term of (\ref{posterior 1}) in ensemble space $\R^L$ we use equation (11) of \cite{Potthast19}, i.e.,\ we have
\beq
q^{(a,\ell)}(\beta) = c_{\ell} e^{-\frac{1}{2}(\beta-e_{\ell})^T (\gamma^{-1} \bI) (\beta-e_{\ell})} 
e^{-\frac{1}{2}[P(y-\bar{y}-\bY \beta)]^{T} \bR^{-1} [P(y - \bar{y} - \bY \beta)]}, 
\;\; \beta \in \R^{L},
\label{posterior 3}
\eeq
for $\ell=1,...,L$, where $P$ is the orthogonal projection onto $\span{\bY}$ with respect to the scalar
product in $\R^m$ weighted by $\bR^{-1}$; 
it is defined in equation (10) of \cite{Potthast19} and Lemma 3.2.3 of \cite{NakaPot} 
to be given by
\beq
P  =  \bY(\bY^{T}\bR^{-1}\bY)^{-1} \bY^{T}\bR^{-1}. 
\eeq
As in (13) - (15) of \cite{Potthast19} the right-hand side of (\ref{posterior 3}) can be 
transformed into 
\beq
q^{(a,\ell)}(\beta) = e^{-\frac{1}{2} (\beta-e_{\ell})^T (\gamma^{-1} \bI) (\beta-e_{\ell})} 
e^{-\frac{1}{2}[C-\beta]^{T} \bA [C-\beta]}, \;\; \ell=1,...,L,
\label{posterior 4}
\eeq
with
\beq
\bA := \bY^{T} \bR^{-1} \bY, \hspace{1cm} C := \bA^{-1} \bY^{T}\bR^{-1}(y-{\myred \oy}). 
\label{ens trans step1}
\eeq
We now carry out (\ref{shift0}) and (\ref{post0}) in ensemble space based on 
(\ref{deriv101}) and (\ref{deriv102a}), leading to the new mean of the posterior 
distribution for the $\ell$-th particle prior distribution 
\beq
\beta^{(a,\ell)} = e_{\ell} + \gamma (\bI + \gamma \bY^{T}\bR^{-1}\bY)^{-1} \bY^{T} \bR^{-1} \bY (C - e_{\ell})
\label{betaa}
\eeq
and the new covariance matrix of this distribution
\beq
{\myred \bG^{(a)}_{ens}} = (\frac{1}{\gamma} \bI + \bY^{T} \bR^{-1} \bY)^{-1} \in \R^{L\times L}
\label{Ga}
\eeq
independent of $\ell$ when {\myblack $\bG = \gamma \bX \bX^{T}$} is independent of $\ell$. This means that we obtain
\beq
q^{(a,\ell)}(\beta) = w_{\ell} e^{-\frac{1}{2} ( \beta - \beta^{(a,\ell)})^{T} \bG^{(a)}_{ens}
(\beta-\beta^{(a,\ell)})}, \;\; \beta \in \R^L
\label{posterior 5}
\eeq
with $\beta^{(a,\ell)}$ given by (\ref{betaa}) and $\bG^{(a)}_{ens}$ given by (\ref{Ga}) for the {\em posterior distribution} of the $\ell$-th particle in ensemble space. 
We denote the term 
\beq
\beta^{(shift,\ell)} :=  \gamma (\bI + \gamma \bY^{T}\bR^{-1}\bY)^{-1} \bY^{T} \bR^{-1} \bY (C - e_{\ell})
\label{shift vector}
\eeq
as the {\em shift vector} for the $\ell$-th particle in ensemble space, i.e., $\beta^{(a,\ell)}=e_{\ell}+\beta^{(shift,\ell)}$ in Eq. \ref{betaa}. The use of the model
error $\gamma \bI$ corresponding to $\gamma \bX \bX^{T}$ for this particle in ensemble space
leads to this shift in the analysis. The shift has important effects:
\begin{enumerate}
\item
it moves the particle towards the observation in ensemble space, 
\item 
by the use of particle uncertainty, it constitutes a further degree of freedom which can be
used for tuning of a real system. 
\end{enumerate}
One of the major advantages and problems at the same time of the LAPF as well as a classical particle filter is that the particles are taken as they are. If the model has some local bias, i.e.,\ if all particles have a similar behaviour and do not fit the observation well, then there is no inherent tool in the classical particle filter or the basic LAPF to move the particles towards the observation - this move is only achieved by selection of the best particles, closest to the observation. By resampling and rejuvenation, effectively the whole ensemble is moved towards the observation. Here, with the introduction of uncertainty of individual particles into the assimilation step, this is already carried out for each individual particle by calculating a posterior mean $\beta^{(a,\ell)}$ in (\ref{betaa}) of the posterior component $q^{(a,\ell)}(\beta)$ given by (\ref{posterior 5}) for the model error prior distribution $q^{(b,\ell)}(x)$ attributed to each particle (\ref{pp}). 

%===============================================================================
%
%
%===============================================================================
\subsection{Putting it all together: the full LMCPF}
\label{lmcpf full}

Here, we now collect all steps to describe the full LMCPF assimilation step and data assimilation cycle. The LMCPF assimilation cycle is run analogously to the LETKF or LAPF assimilation cycle, i.e.,\ we start with some initial ensemble $x^{(a,\ell)}_{0}$ at time $t_0$. Then, for time steps $t_{k}$, $k=1,2,3,...$ we 
%-------------------------------------------------------------------------------
\begin{enumerate}
\item[(1)] carry out a {\em propagation step}, i.e.,\ we run the model forward from time $t_{k-1}$ to $t_{k}$ for each ensemble member, leading to the background ensemble $x^{(b,\ell)}_{k}$ at time $t_k$. 
%-------------------------------------------------------------------------------
\item[(2)] 
Then, at each localization point $\xi$ on a coarser analysis grid $\cG$ we carry out the localized ensemble transform (\ref{ens trans step1}), calculating $C$ and $\bA$. Localization is carried out {\myblack as for the LETKF and LAPF}, i.e.,\ the matrix $\bR$ is weighted depending of the distance of each of its observations to the analysis point. 
%-------------------------------------------------------------------------------
\item[(3)]
We now carry out a classical resampling step following Section 3.d of \cite{Potthast19}. This leads to a matrix 
\begin{equation}
 \breve{\bW}_{i,\ell}=
 \left\{
 \begin{array}{ll}%{@{\kern2.5pt}lL}
   1, & {\rm if} \; R_\ell \in \; (w_{ac_{i-1}}, w_{ac_{i}}], \\
   0, & {\rm otherwise},
 \end{array}
 \right.
\label{Wresampling}
\end{equation}
$i,\ell = 1, ..., L$, draw $r_{\ell}\sim U([0,1])$, set $R_\ell=\ell-1+r_{\ell}$, with accumulated weights $w_{ac}$, 
$w_{ac_0}  = 0, w_{ac_i}  =  w_{ac_{i-1}}+w_{k,i}$ and $\breve{\bW} \in \R^{L\times L}$ defined by {\myblack (\ref{Wresampling})}
with entries one or zero reflecting the choice of particles. {\myblack As for the LETKF and LAPF} this is carried out for each localistion point $\xi$ on a coarser analysis grid $\cG$ to ensure that the weight matrices only change on scales on the order of the localization length scale. Here, we use $\breve{\bW}$ instead of $\breve{\bW(\xi)}$ for brevity.

%-------------------------------------------------------------------------------
\item[(4)]
The posterior matrix {\myred $\bG^{(a)}_{ens}$} given by (\ref{Ga}) and the shift vectors $\beta^{(shift,\ell)}$ given by (\ref{shift vector}) for $\ell=1,...,L$ are calculated for each localization point $\xi$. We define
\beq
\bW^{(shift)} := \Big( \beta^{(shift,1)}, ..., \beta^{(shift,L)} \Big) \;\; \in \R^{L\times L}. 
\eeq
Then, if we want the shift {\myred given by the lth-particle}, we obtain it by the product $\bW^{(shift)} e_{\ell}$. If we have a selection matrix $\breve{\bW}$ for which each column with index $\zeta$, $\zeta=1,...,L$, contains some particle $e_{\ell}$ with $\ell = \ell(\zeta)$, which has been chosen to be the basis for the corresponding new particle, we obtain the shifts for these particles by the product $\bW^{(shift)} \breve{\bW}$. According to the analysis equation (\ref{betaa}) the new coordinates in ensemble space are calculated by
\beq
\Big( \beta^{(a,1)}, ..., \beta^{(a,L)} \Big) = \breve{\bW} + \bW^{(shift)} \breve{\bW}. 
\eeq 
%-------------------------------------------------------------------------------
\item[(5)]
For each particle we now carry out an {\em adaptive Gaussian resampling or rejuvenation} step. The rejuvenation is carried out the same way as described in Section 3.e and 3.f of \cite{Potthast19}, i.e.,\ we first calculate 
\beq
\rho = \frac{{\myred \bd^{T}_{o-b} \bd_{o-b}} - Tr(\bR)}{ Tr(\bH  {\myred \frac{1}{L-1} \bX \bX^{T}} \bH^{T}) }
\eeq
at each localization point, {\myred with the actual ensemble covariance matrix $\frac{1}{L-1} \bX \bX^{T}$ and %{\myblack 
with the observation minus background statistics $\bd_{o-b} = y_k - \bar{y}_k$ where $\bar{y}_k$ denotes the ensemble mean in observation space described in (\ref{ybar}) at time $t_k$.} 
Then we %} 
scale $\rho$ by some function 
\beq
\sigma(\rho) := \left\{
\begin{array}{ll}
c_0, & \rho < \rho^{(0)}, \\
c_0 + (c_1 - c_0) \frac{\rho - \rho^{(0)}}{\rho^{(1)}-\rho^{(0)}}, & 
\rho^{(0)} \leq \rho \leq \rho^{(1)}, \\
c_1, & \rho > \rho^{(1)}, 
\end{array}
\right.
\label{rhop}
\eeq
%%%as chosen in (28) of \cite{Potthast19}, 
{\myblack where the constants $\rho^{(0)}, \rho^{(1)}, c_0, c_1$ are tuning constants.} We note that temporal smoothing is applied to $\rho$ {\myblack as usual for LETKF or LAPF}. Let $\bN \in \R^{L \times L}$ be a matrix with entries drawn from a normal distribution, i.e., each entry is taken from a Gaussian distribution with mean zero and variance 1. This is chosen uniformly for all localization points $\xi$ on the analysis grid $\cG$. Then, the rejuvenation plus shift step is carried out by 
\beq
\bW := \breve{\bW} + \bW^{(shift)}\breve{\bW} + [{\myred \bG^{(a)}_{ens}}]^{\frac{1}{2}} \bN \sigma.
\eeq
Again, we note that $\bW = \bW(\xi)$, $\bW^{(shift)} = \bW^{(shift)}(\xi)$, $\breve{\bW} = \breve{\bW}(\xi)$, $[{\myred \bG^{(a)}_{ens}}]^{\frac{1}{2}} = [{\myred \bG^{(a)}_{ens}}]^{\frac{1}{2}}(\xi)$ and $\sigma = \sigma(\xi)$ are functions of physical space with $\xi \in \cG$ chosen from the analysis grid $\cG$.

%-------------------------------------------------------------------------------
\item[(6)]
The matrices $\bW$ are calculated {\myred at} each analysis point $\xi$ on a coarser global analysis grid $\cG$. 
We now interpolate the matrices onto the full model grid $\cG_{model}$.
\item[(7)]
Finally we calculate the {\em analysis ensemble} (\ref{Xa full})
by
\bea
\bX^{(a,full)} & = & \ox^{(b)} + \bX^{(b)} \bW \label{LMCPF final}
 \\
& = & \ox^{(b)} 
+ \underbrace{ \bX^{(b)} \breve{\bW}}_{class. \; resampling} 
+ \underbrace{ \bX^{(b)} \bW^{(shift)} \breve{\bW}}_{shift} 
+ \underbrace{ \bX^{(b)} [{\myred \bG^{(a)}_{ens}}]^{\frac{1}{2}} \bN \sigma}_{adapt.\; Gauss.\; resampling}
\nonumber 
\eea
\end{enumerate}
%-------------------------------------------------------------------------------
Comparing (\ref{LMCPF final}) with (\ref{LETKF final}) we observe some similarities and some differences. The LETKF does not know the selection reflected by the matrix $\breve{\bW}$, instead it transforms the ensemble by its matrix $\bW$. Both know a shift term, for the LETKF it is given by $w$, for the LMCPF by $\bW^{(shift)}\breve{\bW}$, shifting each particle according to model error (here taken proportional to ensemble spread), where the LETKF shifts according to the full ensemble spread. The LMCPF also takes into account that part of the ensemble spread which is kept during the selection process. Further, it employs adaptive resampling around each remaining shifted particle. {\myred This helps to keep the filter stable and achieve an appropriate uncertainty described by $o-b$ statistics.}

%===============================================================================
%
%
%===============================================================================
\section{Experimental Environment: the Global ICON Model}
\label{environment}

{\myblack \subsection{The ICON Model}}

%%%The core task of our work is to develop non-Gaussian assimilation techniques which can be 
%%%applied to operational data assimilation with its very high-dimensional systems. 
%%%This ranges from basic algorithmical ingredients to 
%%%their application within an operational framework to prove that the techniques can be 
%%%successfully applied. 
%%%We have carried out experiments testing the LMCPF algorithm both for global NWP as well as for 
%%%convective scale assimilation and
%%%forecasting. In this paper, we focus on the global ICON (ICOsahedral Nonhydrostatic) model, i.e.,\
%%%the operational global NWP model of DWD, compare \cite{Zaengl} 
%%%and \cite{Potthast19} for further details on the systems. 
{\myblack We have carried out experiments testing the LMCPF algorithm in the global ICON (ICOsahedral Nonhydrostatic) model, i.e.,\ the operational global NWP model of DWD, compare \cite{Zaengl} and \cite{Potthast19} for further details on the systems.}
ICON is based
on an unstructured grid of triangles generated by subdivision from an initial icosahedron.
The operational resolution is 13 km for the deterministic run and 40 km for the ensembles
both for the data assimilation cycle and the ensemble prediction system (EPS). The upper
air prognostic variables such as wind, humidity, cloud water, cloud ice, temperature, 
snow and precipitation live on 90 terrain-following vertical model levels from the surface
up to 75 km height. In the operational setup, we have 265 million grid points. We also
note that there are further prognostic variables on the surface and on seven soil 
levels, in particular soil temperature and soil water content, as well as snow variables, 
sea ice fraction, ice thickness and ice surface temperature of ICON's integrated sea-ice
model. 

%%%Data assimilation for the deterministic ICON system is based on an {\em Ensemble Variational
%%%Data Assimilation} (EnVAR), where 70\% of the covariance matrix is calculated from 
%%%short-range ensemble forecasts and 30\% are given by the climatological covariance matrix
%%%which is based on the NMC method \citep{Parrish}. The data assimilation for the ensemble
%%%is carried out by a {\em Localized Ensemble Transform Kalman Filter} (LETKF) based on 
%%%\cite{Hunt}. The systems are coupled in the sense that the ensemble is a critical ingredient
%%%of the EnVAR and the quality control of the deterministic part is employed for the 
%%%LETKF. 

{\myblack  The data assimilation for the operational ensemble is carried out by an {\myred LETKF} based on \cite{Hunt}.} We run a data assimilation cycle with an analysis every 3 hours. Forecasts are calculated based
on the analysis for 00 and 12 UTC, with 180 hours forecast lead time. For the operational 
system, forecasts with shorter lead times of 120 hours for 06 and 18 UTC and 30 hours for
03, 09, 15 and 21 UTC are calculated. The ensemble data assimilation cycle 
is run with L=40 members. 

%%%We also note that the system includes modules for {\em snow analysis} (SNOW) every three
%%%hours, {\em sea surface temperature} (SST) analysis and {\em soil moisture analysis} (SMA)
%%%once per day. In the standard mode random perturbations of multiple scales of 100 km and 
%%%300 km for SST and SMA  are added to the ensemble members to guarantee sufficient spread
%%%of surface fields. 

For the experimental setup of our study, we employ a slightly lower horizontal resolution of 52 km for the ensemble and 26 km for the deterministic run {\myblack (in the operational setup a part of the observations quality control is carried out within the framework of the deterministic run, we keep this feature for our particle filter experiments)}. An incremental analysis update with a window of $t \in [-90 \min, 90 \min]$ around the analysis time for starting the model runs is used. The analysis is carried out for temperature, humidity and two horizontal wind components, i.e.,\ for {\em four prognostic variables} per grid point. This leads to $n=6.6 \cdot 10^6$ free variables {\myred at} each ensemble data assimilation step. {\myblack Forecasts are only carried out for 00 and 12 UTC. We employ L=40 members for the experimental runs as well.}
%%%For the experimental setup forecasts are only carried out for 00 and 12 UTC. We employ L=40 members for the experimental runs as well. The modules SNOW, SST and SMA are run in the standard operational mode. This experimental setup is used as a standard building block for the pre-operational testing of new observation systems and improvement of algorithms. 

%===============================================================================
%
%
%===============================================================================
{\myblack \subsection{Comparison in an Operational Framework}}

For testing and developing algorithms in the operational framework, the tuning of basic algorithmic constants is a crucial part. The task of testing in a real-world operational setup is much more intricate than for what is usually done when algorithms are compared in a simulation-only small-scale environment. In particular for new algorithms, the whole {\em model plus assimilation cycled NWP system} needs a retuning and it is difficult to compare one algorithmic layer only within a very complex system with respect to its performance. To compare two algorithms A and B, there are two important points to be taken into account:
\begin{enumerate}
\item[(1)] {\bf Tuning Status of the Methods.}
There might be a {\em raw} or {\em default} version of the algorithms, but when you compare
scores with the task of showing that some algorithm is better than the other, you need to compare
{\em tuned algorithms}. In principle, you have to tune algorithm A to give the best results and then
you have to tune algorithm B to give the best results and then compare the results of tuned A
and tuned B. If A has been tuned for several years, but B is raw, the results give you insight
into the tuning status of A and B, but not necessarily of the algorithms as such! So we have to
be very careful with generic conclusions. 

%%%For example, the operational 
%%%introduction of the KENDA system \citep{Schraff} at DWD in 2017 needed about two years of
%%%working on the LETKF system before it was mature enough and superior to the previous very
%%%well tuned {\em nudging} data assimilation system. 
\item[(2)] {\bf Quality Control of Obervations.}
When you compare two algorithms for assimilation or two models, {\em verification} provides a
variety of scores. But verification with real data needs {\em quality control} of these data, 
since otherwise scores are mainly determined by outliers, and one broken device can make the
whole verification result completely useless. But how is the data quality controlled? Usually we
employ $o-f$ (observation minus first guess) 
statistic and remove observations which are far away from the model first guess. 
This leads to an important point: each algorithm A and B needs to use its own quality control. 
If model biases change between A and B, you will have a different selection of 'good' observations.

But how do you compare two systems which employ different observations? One solution can be to
use observations for comparison which passed both quality controls. A second method is to verify
each algorithm separately and then compare the scores (this is what is done with 
World Meteorological Organization (WMO) score comparisons between global models). 
A third method is to try to use 'independent' observations.
But these also need some quality control, and since they are not linked to any of the forecasting
systems, it is unclear in what way their use in verification helps to judge a particular 
algorithm or to compare two algorithms. 
\end{enumerate}

For our experiments, we compare the LMCPF with the LAPF and the LETKF. The LETKF has a relatively
advanced tuning status. LAPF has been mildly tuned and the LMCPF is relatively new. We carried
out several tuning steps to try to make LMCPF and LETKF comparable. Further, we employ quality
control for the observations in each system separately. Verification of the $o-f$ statistics 
is based on each system independently. Here, one important performance measure is the number
of observations which passes the quality control. If these number is larger for B than for A, 
we can conclude that the system fits better to the observations, which is a good indicator for the 
quality of a short-range
forecast. For comparison of forecasts the joint set of observations is used, those which pass
both the quality control of algorithm A and algorithm B. 

%===============================================================================
%
%
%===============================================================================
\section{Numerical Results}
\label{numerics}

The goal of this numerical part is, {\em firstly}, to investigate the relationship between the observation vector mapped into ensemble space and the ensemble distribution. {\em Secondly}, we show since the LMCPF moves particles based on the Gaussian uncertainty of individual particles, it bridges the gap between forecast ensemble and observations. Furthermore we study its distribution. The {\em third} part shows results of {\em observation - first guess (o-f)} statistics for the LMCPF with different choices for $\kappa>0$ compared to the LETKF and the LAPF. {\em Fourthly}, we investigate the evolution of ensemble spread with different parameter settings. In the last part we demonstrate the feasiblity of the LMCPF as a method for atmospheric analysis and subsequent forecasting in a very high-dimensional operational framework, demonstrating that it stably runs for a month of global atmospheric analysis and forecasting. 

%-------------------------------------------------------------------------------
%
%-------------------------------------------------------------------------------

\begin{figure}
\centering
\includegraphics[width=15cm]{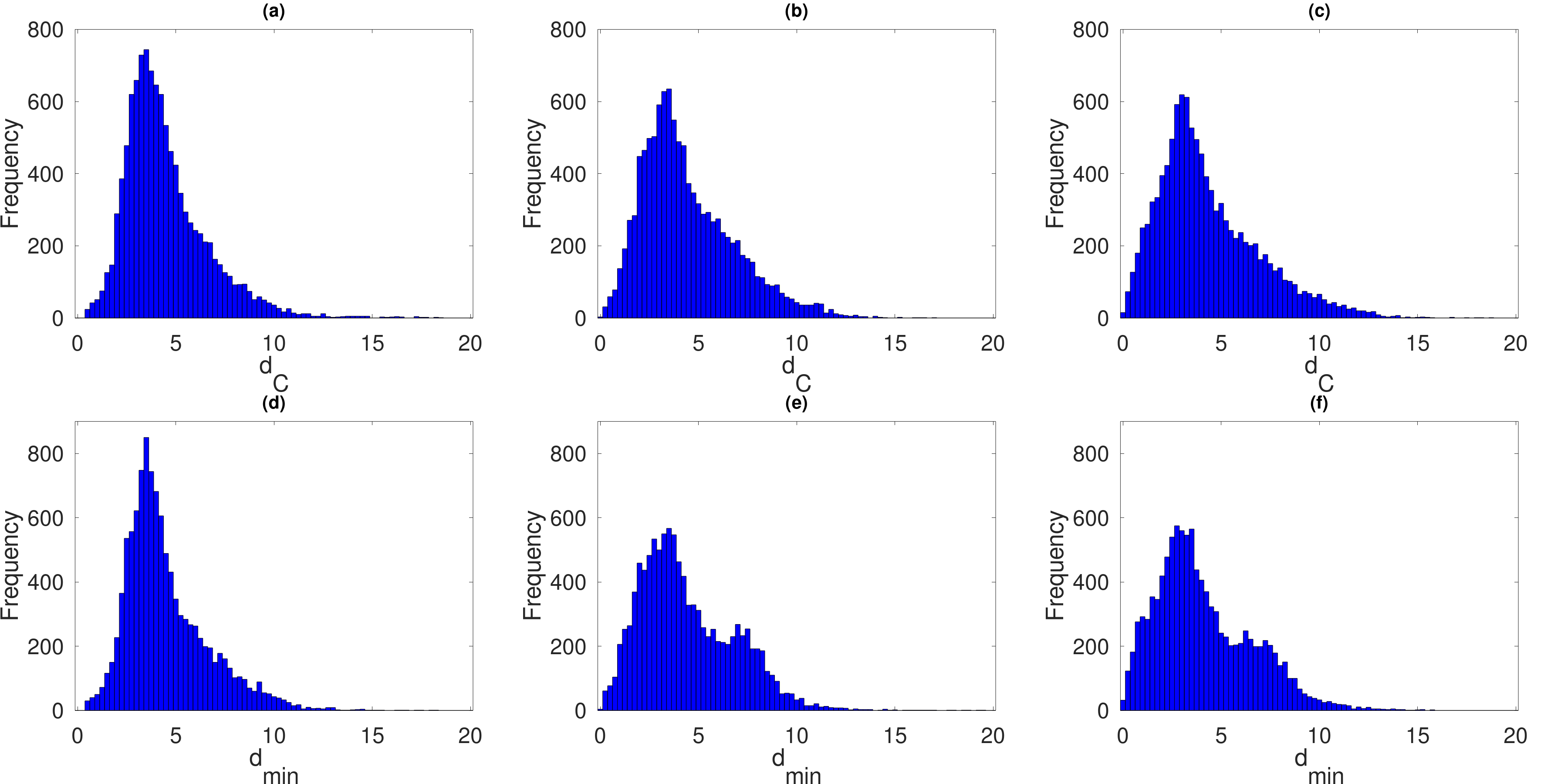}
\caption[We show global histograms of $d_C$ and $d_{min}$ for three different pressure levels.]{
We show global histograms of $d_C$ and $d_{min}$ defined in (\ref{d_C def}) and (\ref{d_min def}) for three different pressure levels:
500 hPa in (a) and (d), 850 hPa in (b) and (e) and 1000 hPa in (c) and (f),
with $d_C$ in (a)-(c) and $d_{min}$ in (d)-(f). 
Shown are statistics for the LMCPF with $\kappa=25$ for May 6th, 0 UTC.}
\label{dCdmin}
\end{figure}

%===============================================================================
%
%
%===============================================================================
{\myblack \subsection{Distributions of Observations in Ensemble Space}}

In a first step, we study (a) the distance between the observation and the ensemble mean and (b) the minimum distance between the observation and the ensemble members. In ensemble space, for distance calculations an appropriate metric needs to be used. Recall that $\R^m$ with dimension $m$ is the observation space and $\R^L$ with dimension $L$ the ensemble space. Given a vector $\beta\in\R^{L}$ in ensemble space, the distance corresponding to the physical norm $\norm{\cdot}_{R^{-1}}$ in observation space, which is relevant to the weight calculation of the particle filter, is calculated by
\bea
\norm{\bY\beta}^2_{R^{-1}} & = &  \langle \bY\beta,\bY\beta \rangle_{R^{-1}} \nonumber \\ 
& = & \langle \bY\beta,\bR^{-1}\bY\beta \rangle \nonumber \\
& = & (\bY\beta)^T\bR^{-1}\bY\beta \nonumber  \\
& = & \beta^T(\bY^T\bR^{-1}\bY)\beta \nonumber \\
& = & \langle \beta,\bA\beta \rangle \nonumber \\
& = & ||\beta||^2_\bA
\label{length}
\eea
where $\langle\cdot,\cdot\rangle$ denotes the standard $L^2$-scalar product in $\R^m$ or $\R^L$, respectively. The notation $\langle \cdot,\cdot\rangle_{\bD}$ with some positive definite matrix $\bD$ denotes the weighted scalar product $\langle \cdot, \bD \; \cdot \rangle$ and $\norm{\cdot}_{\bD} = \langle \cdot,\cdot \rangle_{\bD}$, here with either $\bR^{-1}$ in $\R^m$ or $\bA$ in $\R^L$. {\myblack Note that for $A$ to be positive definite we need $L\leq m$.}  

The matrix $\bA$ including the standard LETKF localization in observation
space has been integrated into the data assimilation coding environment. Here, 
we show results from an LMCPF one month experiment studying one assimilation step
at 0 UTC of May 6, 2016. The cycle has been started May 1, such that the results
illustrate a situation where the spin-up period is over and LMCPF spread has
reached a steady state (compare Figure \ref{spread}). 

%-------------------------------------------------------------------------------
%
%-------------------------------------------------------------------------------
\begin{figure}[ht]
\centering
\includegraphics[width=15cm]{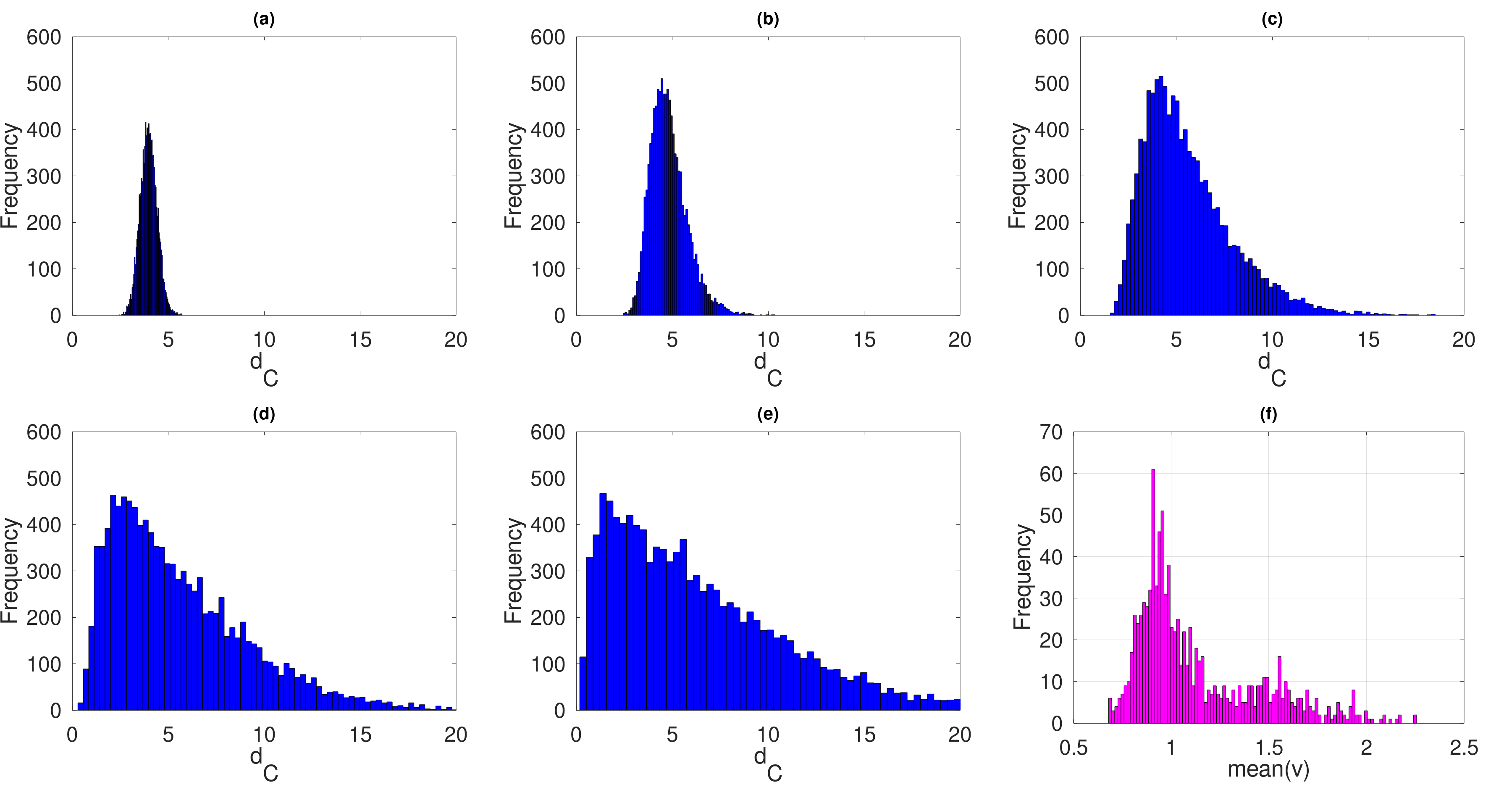}
\caption[We show simulations of distributions of random draws with different mixtures of variances. A histogram of the fit of exponents $\nu$ for a selection of 1000 points is shown in (f).]{We show simulations of distributions of random draws in an $L=40$ dimensional space, with different mixtures of variances given by formula (\ref{std_sim}), here with $\eta \in (4, 15, 30, 40, 50)$ and $\nu \in (0, 0.5, 1, 2, 3)$ in (a) to (e). A histogram of the fit of exponents $\nu$ as in (\ref{std_sim}) to the eigenvalue decay of the matrices $\bA$ for a selection of 1000 points is shown in (f). The fit is obtained from the mean of exponents derived from formula (\ref{nu}).}
\label{simFig}
\end{figure}

At each analysis grid point $\xi$ of some coarse global analysis grid $\cG$ we have a matrix $\bA$ (see Eq. (\ref{ens trans step1})), $L=40$ ensemble members and one projected observation vector $C\in\R^L$ (see Eq. (\ref{ens trans step1})). This leads to a total of $N_{\omega} = 10890$ samples $\omega$ {\myblack numbering the analysis grid points in a given height layer,} e.g.\ for 850 hPa. The distance of the observations to the ensemble mean is given by
\begin{equation}
\label{d_C def}
d_{C}(\omega) := \norm{C(\omega)}_{\bA(\omega)},  
\end{equation}
{\myblack where the metric $A$ is chosen to be consistent with (\ref{posterior 4}).}
The minimal distance of the observations vector to the ensemble members
is given by 
\begin{equation}
\label{d_min def}
d_{min}(\omega) := \min_{j=1,...,L} \norm{C(\omega) - e_{j}}_{\bA(\omega)}, 
\end{equation}
with $\omega=1,...,N_{\omega}$, {\myblack where we employed (\ref{length}) and where we note that in ensemble space the ensemble members $x^{(b,j)}-\overline{x}$ are given by the standard unit normal vectors $e_{j}$, $j=1,...,L$.}  

To analyse the role of moving particles towards the observation in ensemble space, in Figure \ref{dCdmin} we show global histograms for $d_C$ and $d_{min}$ for three height levels of approximately 500 hPa, 850 hPa and 1000 hPa. {\myred When the distribution of both $d_C$ and $d_{min}$ are similar, i.e.\ the distribution of the minimal distance of the observation to the ensemble members and the distribution of the distance of observations to the ensemble mean are comparable, it indicates that we have a well-balanced system.} To understand the particular form of the distributions, we compare it with simulations of random draws of a Gaussian distribution in a 40 dimensional space shown in Figure \ref{simFig}. When you draw from a Gaussian with mean zero and standard deviation $\sigma = 4$, we obtain Figure \ref{simFig} (a). The behaviour of the histograms of the norms of the points drawn changes significantly if we consider mixtures with different variances in different space directions. Figure \ref{simFig} (a)-(e) shows different distributions with variances given by 
\begin{equation}
\label{std_sim}
\sigma_j = \frac{\eta}{j^{\nu}}, \;\; j=1,...,L
\end{equation}
where the constant $\eta \in (4, 15, 30, 40, 50)$ has been chosen to achieve a maximum around $4$ and different decay exponents $\nu \in (0, 0.5, 1, 2, 3)$ have been tested. The distributions of Figure \ref{dCdmin} correspond to a decay exponent between $\nu = 1$ and $\nu = 2$. How much is this reflected by the eigenvalue distributions for the matrices $\bA$? We have carried out a fit to the eigenvalue decay of $\bA$ for a selection of analysis points. {\myblack The constant $\eta$ is obtained by using $j=1$, which leads to $\sigma_1 = \eta$. Taking the logarithm on both sides now yields
\begin{equation}
\nu \log(j) = \log(\eta) - \log(\sigma_j), \;\; j=2,...,L. 
\label{nu}
\end{equation}
A fit of $\nu$ can be obtained for example by division through $\log(j)$ and taking the mean of the remaining right-hand side. The distribution of the resulting exponents is displayed in Figure \ref{simFig} (f).} The results find exponents between 0.7 and 2.2. {\myred The corresponding distributions are those shown in Figure \ref{simFig}(c) and (d), which are quite close to the distributions of $d_C$ found in the empirical particle-filter generated NWP ensemble Figure \ref{dCdmin}.} 

%===============================================================================
%
%
%===============================================================================
\subsection{The Move of Particles}

\begin{figure}[ht]
\centering
\includegraphics[width=12cm]{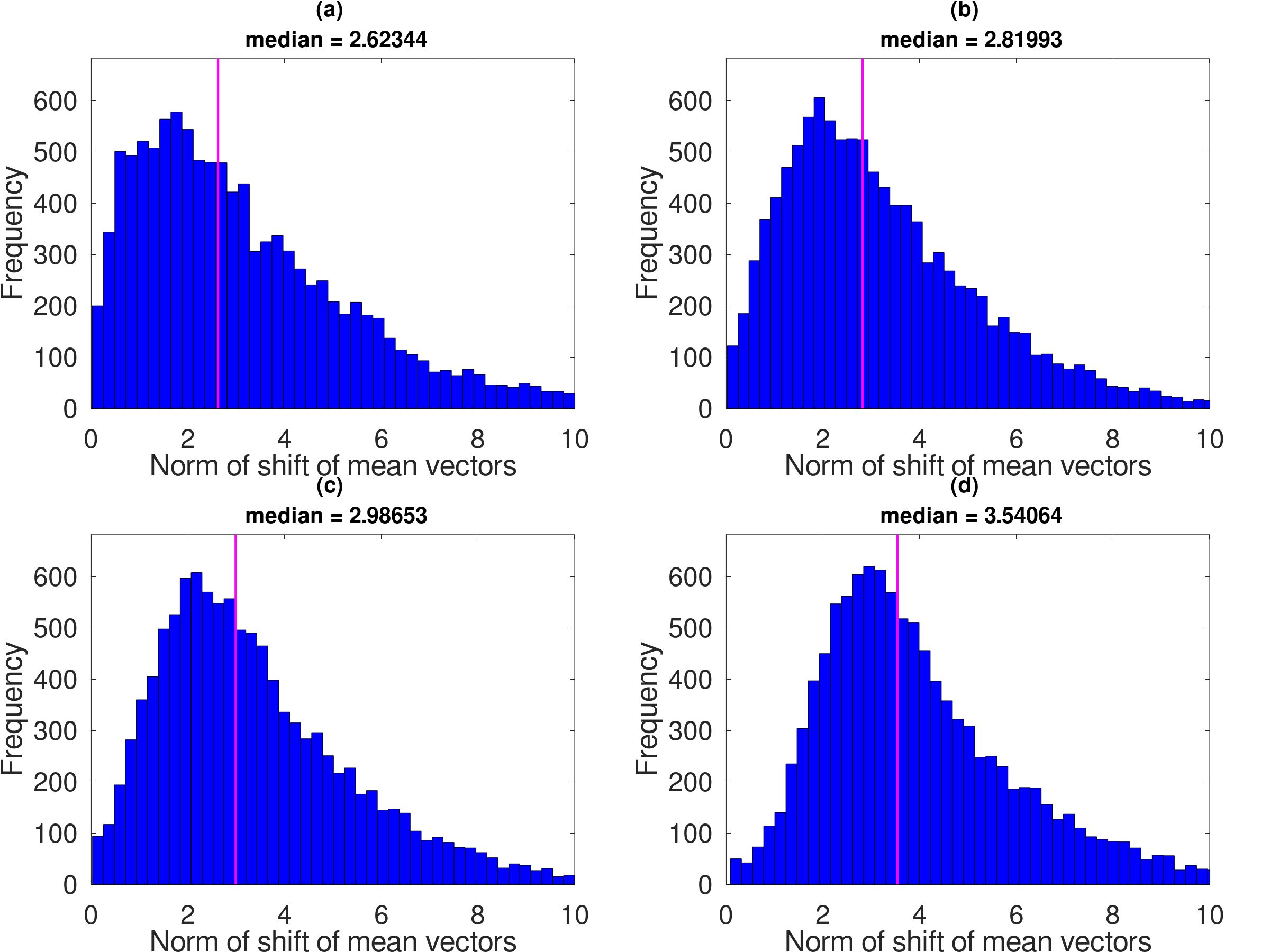}
\caption[We show global histograms of the norm of the mean ensemble shift at pressure level at 500hP]{  
We show global histograms of the norm of the mean ensemble shift at pressure level at 500hPa. On the x-axis we show the norm of shift of mean vectors in ensemble space and on the y-axis we show the frequency. We display the histogram for (a) the LAPF, (b) the LMCPF with $\kappa = 1$, (c) the LMCPF with $\kappa = 2.5$ and (d) shows the LMCPF with $\kappa = 25$. The pink line displays the median, which is also shown on the top of each plot. Shown are the statistics for May 6th, 0 UTC. 
\label{hists}}
\end{figure}

In a second step, we want to investigate the capability of the LMCPF to move particles towards the observation by testing different choices of $\kappa>0$ given by (\ref{kappa def}). In Figure \ref{hists} we compare histograms of the norm of the mean ensemble shift in ensemble space for pressure level 500 hPa, determined for May 6th, 0 UTC. The four histograms show the statistics for the three filters in different settings: a) LAPF, b) LMCPF with $\kappa = 1$, c) LMCPF with $\kappa = 2.5$ and d) LMCPF with $\kappa = 25$. 

%-------------------------------------------------------------------------------
%
%-------------------------------------------------------------------------------

There are two effects seen in Figure \ref{hists}. First, we see the distribution of average shifts or moves of the ensemble mean generated by the LAPF and the LMCPF with three different choices $\kappa$ controlling the size of the uncertainty used for each particle. The mean shift increases if the uncertainty increases, i.e.,\ from $\kappa=1$ to $\kappa =2.5$ and $\kappa=25$. To develop an understanding of the relative size of this shift let us look at the one-dimensional version of formula (\ref{shift vector}) given by
\begin{equation}
\label{s factor}
s(\kappa) = \frac{\kappa b}{r + \kappa b}, 
\end{equation}
with background variance $b$ and observation error variance $r$, reflecting the size of the particle move. When we, for example, choose $r = 4$ and $b = 16$, as we would get with {\myred typical values for the} error of $2\ \frac{m}{s}$ for wind measurements and an ensemble standard deviation of $4\ \frac{m}{s}$, and then study $\kappa \in (1, 2.5, 10, 25)$, we obtain factors of size $s(\kappa) \in (0.8, 0.9, 0.97, 0.99)$. If the observation has a distance of $3.6$ to the ensemble mean, {\myred as seen in Figure \ref{dCdmin},} this would make the means observed in Figure \ref{hists} plausible. For small $\kappa=1$ here the particle move is 0.8 times the innovation, for large $\kappa=25$ it is 0.99 times the innovation $y-H(x^{(b)})$. In Figure \ref{hists} we observe this behaviour with the median of the ensemble increments being {\myblack ${\tt median}=2.62$} in (a) to {\myblack ${\tt median}=3.54$} in (d). 

%-------------------------------------------------------------------------------
%
%-------------------------------------------------------------------------------
\begin{figure}
\centering
\includegraphics[width=15cm]{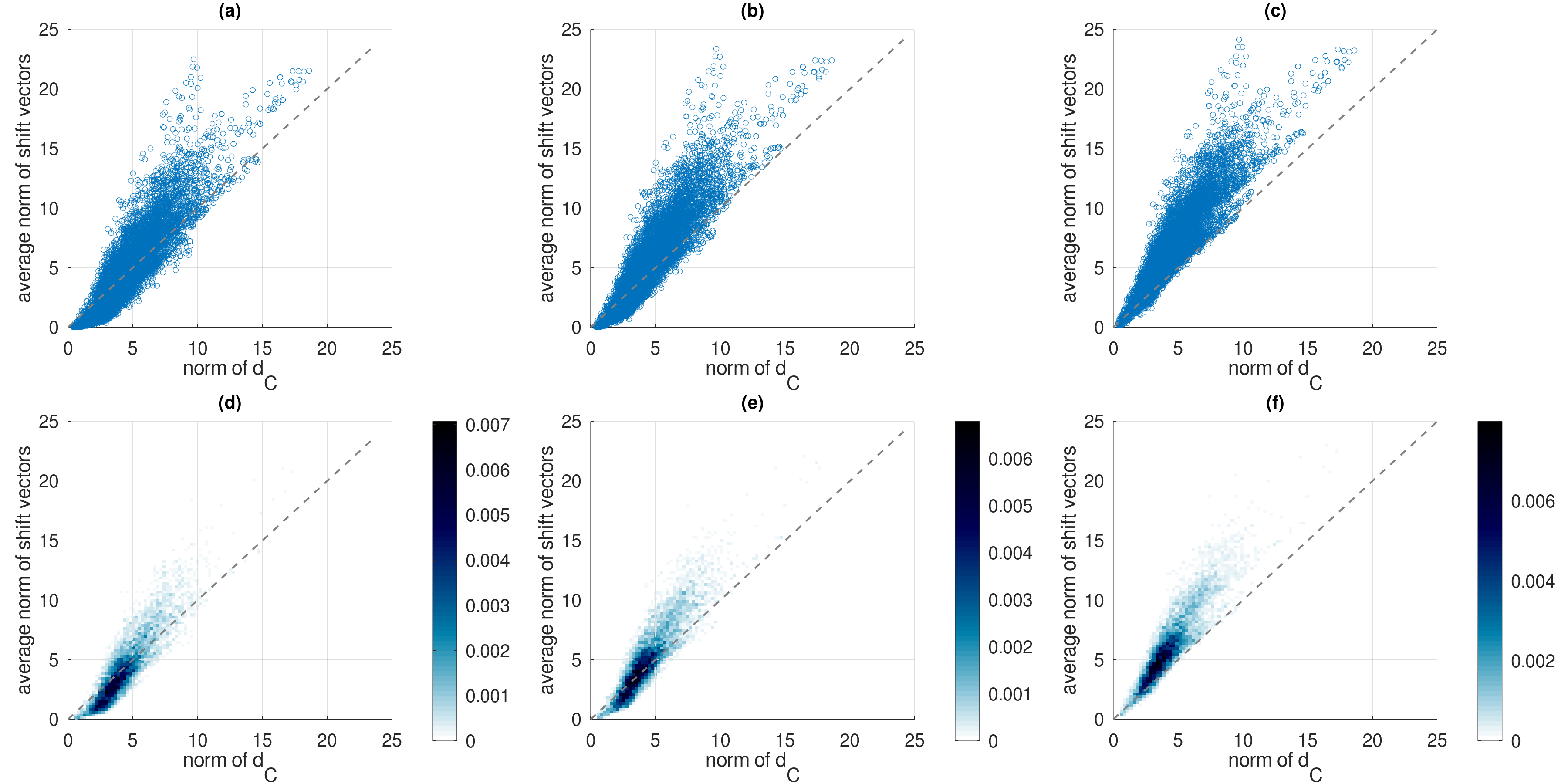}
\caption[We show scatter and density plots of the average particle move versus the distance of the observation vector to the ensemble mean.]{
We show scatter and density plots of the average particle move versus the distance of the observation vector to the ensemble mean, all for the pressure level 500 hPa in ensemble space. On the x-axis we can see the norm of the observation distance to ensemble mean and on the y-axis we show the average size of the corresponding particle move. We display statistics for the LMCPF with different particle uncertainty, for each setting a scatter plot and a density plot which shows high density of points in a better way. (a) and (d) show the statistics for $\kappa = 1$, (b) and (e) for $\kappa = 2.5$ and (c) and (f) for $\kappa = 25$, all for May 6th, 0 UTC.
}
\label{scatter}
\end{figure}

%-------------------------------------------------------------------------------
%
%-------------------------------------------------------------------------------
\bigskip
As a final step of this part, we want to investigate not only the overall distribution of the particle moves, but relate the size of the average particle move to the distance of the observation to the ensemble mean. 
Figure \ref{scatter} shows scatter and density plots for the LMCPF with different particle uncertainty. We employ the same values for $\kappa$ as in Figure \ref{hists}, (a) and {\myred (d)} with $\kappa=1$, {\myred (b) and (e)} with $\kappa=2.5$, {\myred (c)} and (f) show results for $\kappa=25$. Displayed are statistics for the average particle move vs. the difference of the observation vectors from the ensemble mean, all for the pressure level at 500 hPa.

The results of Figure \ref{scatter} show that clearly the move of the
particles is related to the necessary correction as given by the distance of the observation to the individual particle. There is a clear correlation of the average move to the observation discrepancy with respect to the ensemble
mean. If we would investigate each particle individually in one dimension, all points would be on a straight line with slope given by (\ref{s factor}). The situation in a high-dimensional space with non-homogeneous metric is more complicated as reflected by Figure \ref{scatter}. The figure confirms that the method is working as designed. 

%===============================================================================
%
%
%===============================================================================
\subsection{Assimilation Cycle Quality Assessment of the LMCPF}

Here, {\myblack studying} standard global atmospheric scores for the analysis cycle we investigate the quality of the LMCPF by testing different choices of $\kappa>0$, investigate the interaction effects between particle uncertainty, ensemble spread and adaptive spread control and compare it to the way the LETKF moves the mean of the ensemble. For this aims we show two figures. 

%-------------------------------------------------------------------------------
% 
%-------------------------------------------------------------------------------
\begin{figure}[ht]
\centering
\includegraphics[height=8cm]{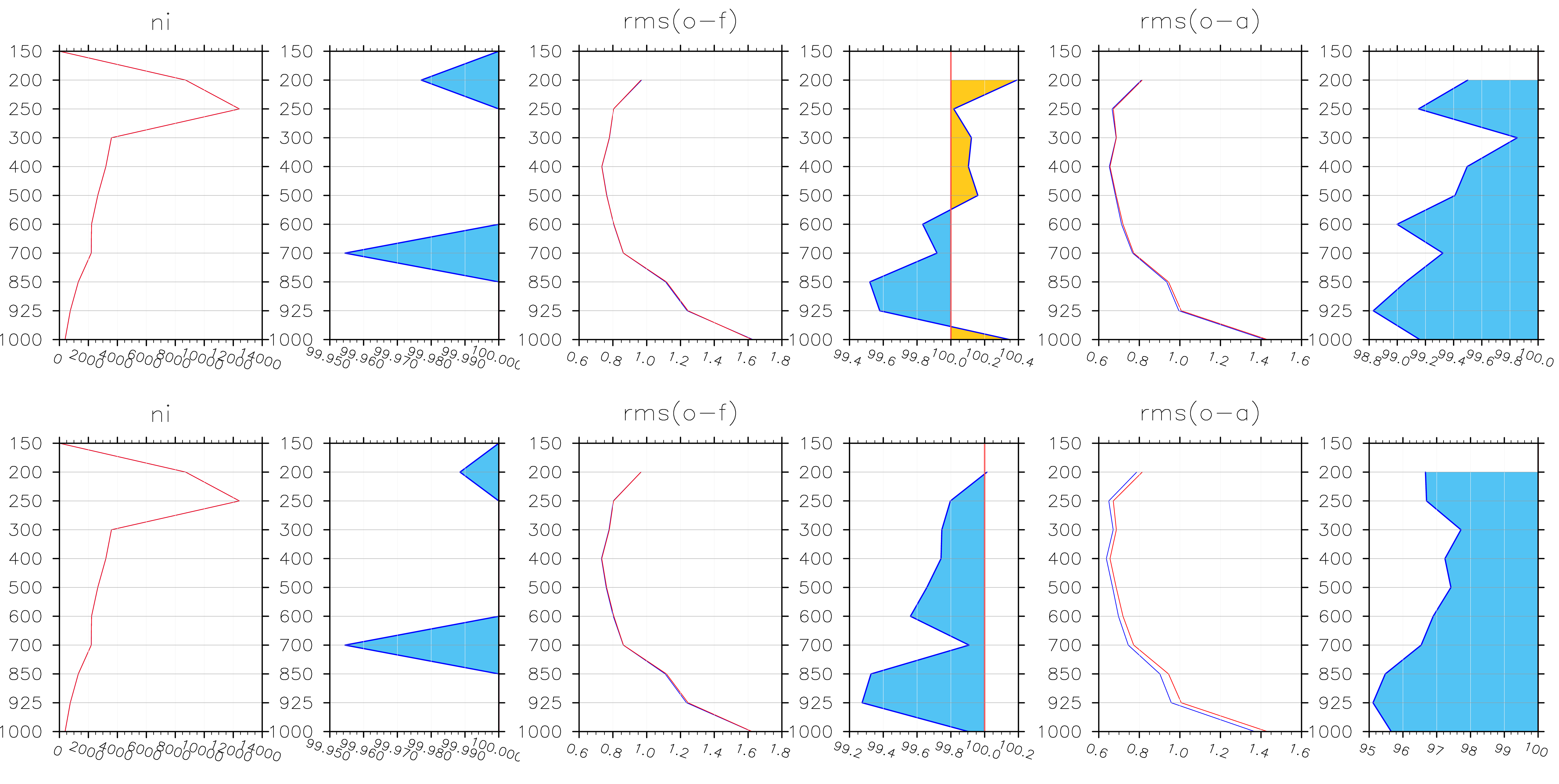}
\caption[We show the observation verification of {\em upper air temperature} measured by airplanes, in particular the first guess and analysis scores]{ 
We show the observation verification of {\em upper air temperature} measured by airplanes, in particular the first guess and analysis scores. The three columns show the number of observations which passed quality control, the RMSE for $o-f$ statistics and the RMSE for $o-a$ statistics for two different choices of particle uncertainty. We display results for one global assimilation step at 20160506 03 UTC. In both cases the comparison of the LETKF (red line) with the LMCPF (blue line) is shown. In the first row we choose $\kappa=1$, for the second row $\kappa=2.5$} 
\label{obs_err1}
\end{figure}
%
%-------------------------------------------------------------------------------
% 
%-------------------------------------------------------------------------------
Figure \ref{obs_err1} shows the functionality of the LMCPF by a display of the analysis and the first guess errors for upper air temperature for an ICON assimilation step, comparing the LETKF and the LMCPF with two different values of particle uncertainty controlled by $\gamma$ or $\kappa$, respectively. Here, we show statistics for the LMCPF (blue line) with $\kappa=1$ in row one and in row two we show the LMCPF (blue line) with $\kappa = 2.5$, both rows are showing the difference to the LETKF (red line). The left panel shows the number of observations which passed quality control, the middle panel shows the root mean square error (RMSE) of observation minus first guess statistics ($o-f$) (also known as observation - background ($o-b$) statistics) and the right panel shows the RMSE for observations minus analysis statistics ($o-a$). The blueish shading shows areas with lower values for the LMCPF in comparison to the LETKF.

It can clearly be seen that with respect to $o-f$ scores the LMCPF is able to outperform the LETKF in case studies with one assimilation step {\myred when an appropriate size of the uncertainty of each particle, here given by the size of $\kappa$, is found}. In the large model error case we observe up to 5\% improvement for the $o-a$ RMSE and 0.8\% for the $o-f$ RMSE. These RMS errors for $o-a$ and $o-f$ are higher for the LMCPF with smaller particle uncertainty, {\myred but even the LMCPF with smaller particle uncertainty is able to outperform the LETKF in $o-f$ statistics for heights between 600 hPa and 925 hPa.}

The numerical experiments prove that the particle uncertainty enables the LMCPF to move the background ensemble towards the observation in a way comparable to or even better than the LETKF. This effect remains active during model propagation and can also be observed for the first guess statistics and for forecasts with short lead times. Here, the LMCPF is able to outperform the operational version of the LETKF.

%-------------------------------------------------------------------------------
% 
%-------------------------------------------------------------------------------
\begin{figure}
\centering
\includegraphics[height=8cm]{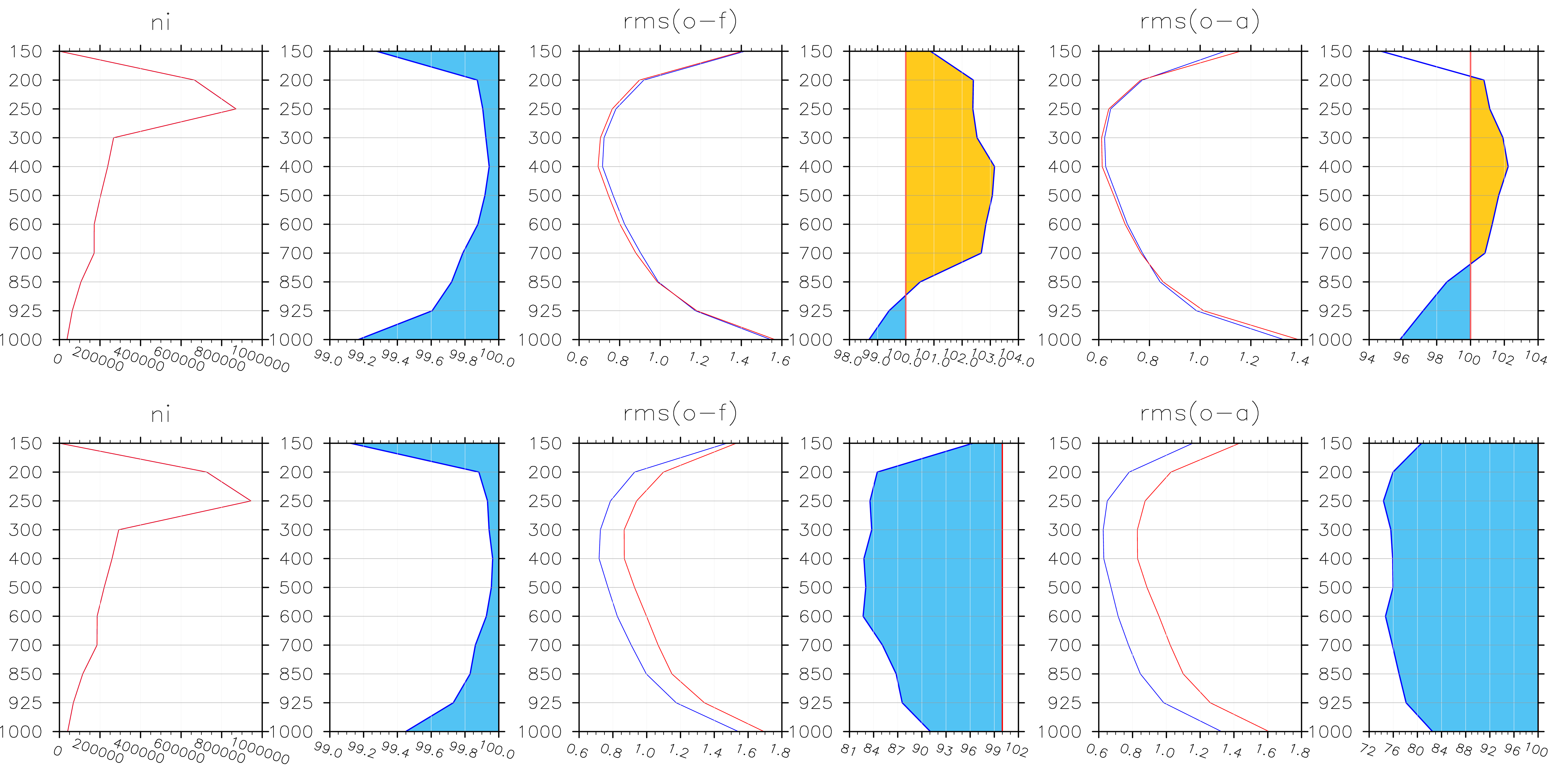}
\caption[Again, we show some observation verification statistics for {\em upper air temperature} measured by airplanes, for three different experiments carried out for the period of May 2016]{Again, we show some observation verification statistics for {\em upper air temperature} measured by airplanes. We show the {\myred same statistics as in Figure \ref{obs_err1} but} for three different experiments carried out for the period of May 2016. In the upper row the comparison between LETKF (red line) and LMCPF (blue line) is shown, in the lower row we show the comparison between LAPF (red line) and LMCPF (blue line).}
\label{obs_err2}
\end{figure}

%-------------------------------------------------------------------------------
% 
%-------------------------------------------------------------------------------
In Figure \ref{obs_err2} we show a comparison of analysis cycle verification for a full one month period of LMCPF, LAPF and LETKF experiments. {\myred The columns are showing the same statistics as in Figure \ref{obs_err1}.} The first row in Figure \ref{obs_err2} shows the differences between LETKF (red line) and LMCPF with $\kappa=2.5$ (blue line) for a full month of cycling. The second row shows the comparison of LAPF (red line) and LMCPF with $\kappa=2.5$ (blue line) for the same period of time. Again, the blueish shading indicates lower numbers or RMSE values for the experiment (LMCPF), the yellowish shading  indicates lower values for the reference (LETKF resp. LAPF).

Row one shows that the LMCPF with particle uncertainty given by $\kappa=2.5$ can outperform the LETKF in regions below a height of 850 hPa and for short lead times -- a very important region and time scale for practical applications. Here the LMCPF is up to 1.5\% better than the LETKF for the $o-f$ statistics. In this experiment, for higher levels in the atmosphere the $o-a$ and $o-f$ statistics of the LMCPF are up to 3.5\% worse than the LETKF. The amount of data which passes quality control is quite similar for all methods under consideration, however, at some levels we loose up to 0.9\% of observations in comparison with the LETKF. This is an effect of quality control based on the ensemble spread - a smaller ensemble spread as we observe for the particle filter leads to less observations passing quality control. In the second row of Figure \ref{obs_err2} we show the statistics of LAPF \citep{Potthast19} vs. LMCPF. Here we can clearly see that the LMCPF shows much better upper air scores than the LAPF. It clearly shows the importance to allow a movement of particles towards the observations by using particle uncertainty.

Overall we conclude that with respect to the verification of the analysis cycle the LMCPF with particle uncertainty given by $\kappa = 2.5$ is comparable to the LETKF, with some levels to be better, some to be worse, overall differences mostly below 3\%. The upper air verification for the analysis cycle of the LMCPF in operational setup is more than 10\% better than for the LAPF.  

%===============================================================================
%
%
%===============================================================================
\subsection{The Evolution of the Ensemble Spread}

\begin{figure}
\centering
\includegraphics[height=11.5cm]{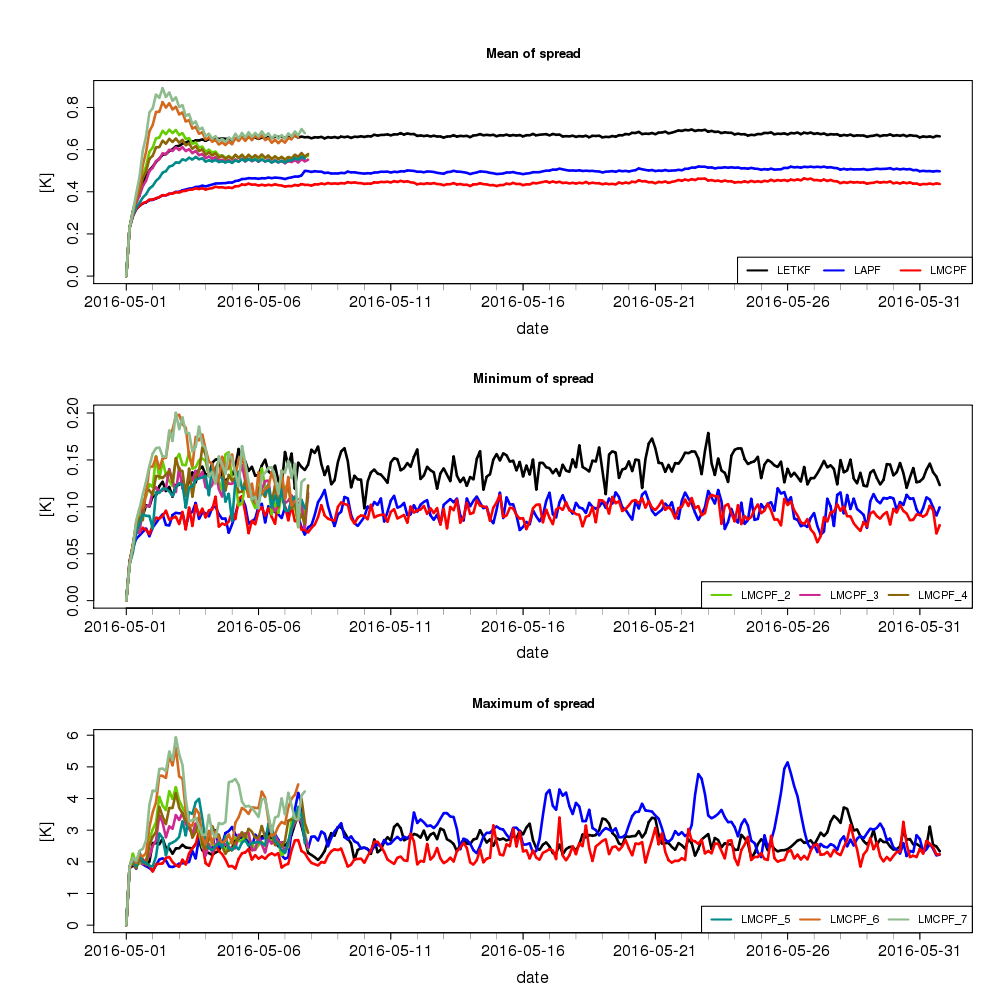}
\caption[The evolution of the ensemble spread is shown for three filters and six different parameter choices for the LMCPF.]{The evolution of the ensemble spread is shown for three filters and six different parameter choices for the LMCPF for a time period of both one month (LETKF - black, LAPF - blue, LMCPF - red) and for one week for different parameter choices for the LMCPF (see Table \ref{Tab_spread}). The x-axis shows the period in one day steps. The y-axis shows the upper air temperature at ICON model level 64 ($\approx$ 500 hPa) in Kelvin. The first row shows the mean of the spread, the second row the minimum and the third row the maximum.}
\label{spread}
\end{figure}

It is an important evaluation step to investigate the stability of the LMCPF for global NWP over longer periods of time. To this end, we have run a period of one month. We compare the particle spread evolution of the LMCPF, the LAPF and LETKF in Figure \ref{spread}. All experiments were started with an ensemble which consists of 40 identical copies of the particles, i.e.,\ with an ensemble in degenerate state. Thus, here the tests also evaluate the capability of the whole system to resolve degeneracy and return to an ensemble with reasonable stable spread. 
 
%-------------------------------------------------------------------------------
% 
%-------------------------------------------------------------------------------

%-------------------------------------------------------------------------------
% 
%-------------------------------------------------------------------------------
In a sequence of experiments we have tested the ability of the LMCPF to reach 
and maintain a particular ensemble spread using {\myred a combination of the choice of $\kappa$ with a} posterior covariance inflation 
\begin{equation}
{\myred \tilde{\bG}^{(a)}_{ens} = \kappa_{post} \bG^{(a)}_{ens}}
\end{equation}
for each particle with $\tilde{\bG}^{(a)}_{ens}$ replacing $\bG^{(a)}_{ens}$ in equation (\ref{LMCPF final}), which is used to generate the analysis ensemble by random draws. We also note that for the random draw of equation (\ref{rhop}) we employed bounds given by $c_0$ and $c_1$. The parameter combinations chosen for six different experiments over one week are compiled into Table \ref{Tab_spread}. The corresponding spread evolution is visualized in Figure \ref{spread}. The results show that, starting with an initial ensemble of identical particles, after some spin-up phase of 2-3 days all particle filters reach their particular spread level and keep it stable over a longer period of time. We carried out selected longer term studies comparing the behaviour of the LMCPF (red), the LAPF (blue) and the LETKF (black) over a period of one month. 

%-------------------------------------------------------------------------------
% 
%-------------------------------------------------------------------------------
\begin{table}
\begin{center}
\begin{tabular}{|p{2cm}|p{1.5cm}|p{1.5cm}|p{1.5cm}|p{1.5cm}|p{1.5cm}|}
\hline
Exp No. & $\kappa$ & $\kappa_{post}$ & $c_1$ & $\rho^{(1)}$ \\ 
\hline
2 & 0.5 & 5 & 0.5 & 1.5 \\
\hline
3 & 0.5 & 3 & 0.5 & 1.5 \\
\hline
4 & 0.3 & 5 & 0.5 & 1.5 \\
\hline
5 & 1 & 1 & 0.3 & 3.0 \\
\hline
6 & 0.5 & 3 & 0.5 & 3.0 \\
\hline
7 & 0.3 & 5 & 0.5 & 3.0 \\
\hline
\end{tabular}
\end{center}
\caption[Parameter choices for the six one week experiments of Figure \ref{spread}.]{Parameter choices for the six one week experiments of Figure \ref{spread}. Further, we used $c_0 = 0.02$ and $\rho^{(0)} = 1.0$ for all experiments. 
}
\label{Tab_spread}
\end{table}

%-------------------------------------------------------------------------------
% 
%-------------------------------------------------------------------------------
The control of the ensemble spread is a delicate topic. A larger ensemble spread does not necessarily lead to better forecast scores, measured by RMSE (Skill) of the ensemble mean or its standard deviation (SD), defined as the RMSE after the bias has been subtracted. With the ability to control separately the strength of the adaptive resampling and the ability of the filter to pull the particles towards the observations, we have independent parameters at hand to adapt the approximations to a real-world situation. At the same time, the way the assimilation step of the LMCPF pulls the ensemble to the observations is based on both the size of the {\em particle uncertainty}, which itself is depending on the {\em ensemble spread}, and within the cycled environment on the adaptive resampling. Of course, it would be desirable to develop tools to estimate the real uncertainty adequate for each particle, and to keep all parts of the system consistent. We expect this to lead to much further research and discussions, which are beyond the scope of this work.  

%===============================================================================
%
%
%===============================================================================
\subsection{Forecast Quality of the LETKF and LMCPF Experiments}
\label{experiments}

\begin{figure}
\centering
\includegraphics[height=12cm]{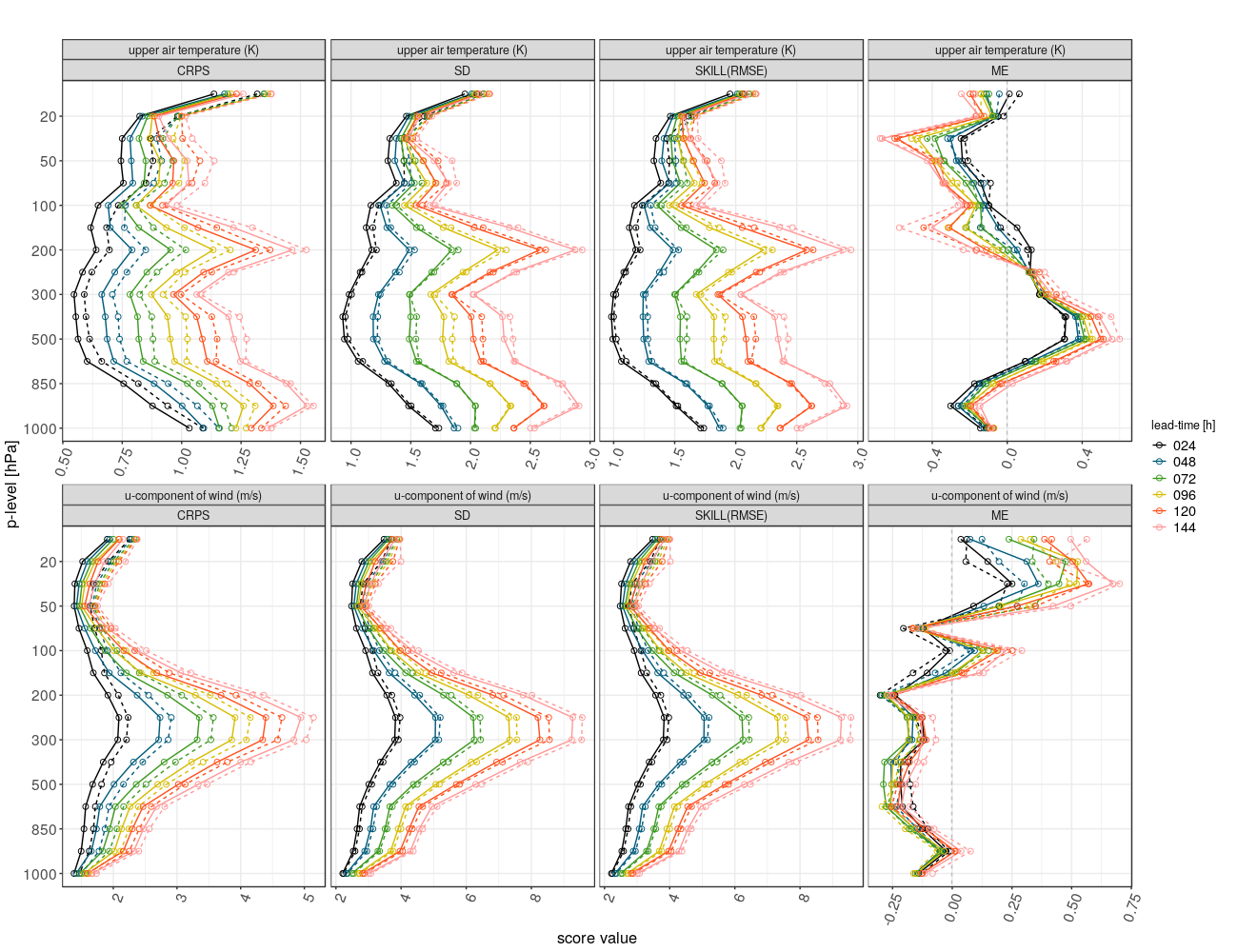}
\caption[Forecast scores for the LMCPF and the LETKF.]{{\myblack We display forecast scores for the LMCPF (dashed) and the LETKF (bold lines). Shown are the continuous rank probability score (CRPS), the standard deviation (SD), the RMSE and the mean (ME). First row shows the upper air temperature, the second row shows the u-component of wind measured by radiosondes. The colors indicate the different lead times from one day to 7 days.}}
\label{veri1}
\end{figure}

\begin{figure}
\centering
\includegraphics[height=12cm]{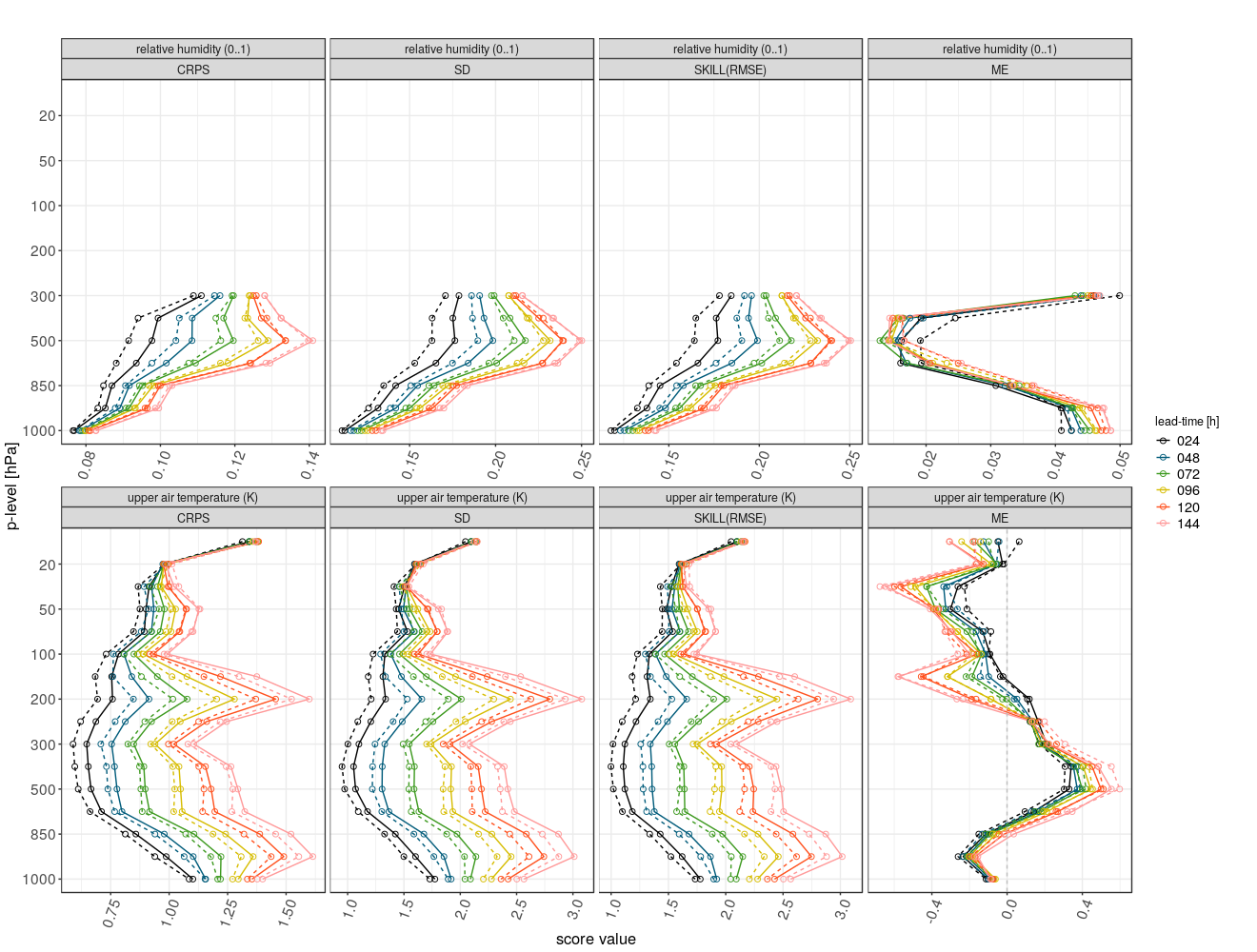}
\caption[Same as in Figure \ref{veri1}, but just for relative humidity and upper air temperature for LMCPF and the LAPF.]{{\myblack Similar to Figure \ref{veri1}, but} for relative humidity and upper air temperature 
for LMCPF (dashed) and the LAPF (bold lines).}
\label{veri2}
\end{figure}

As the last part of the numerical results, we study the quality of longer {\em forecasts} based on the analysis cycle of the LMCPF with $\kappa=2.5$ and compare it to the LETKF based forecasts in Figure \ref{veri1} and to forecasts based on the LAPF analysis cycle in Figure \ref{veri2}. For this purpose, forecasts were run twice a day at 00\ UTC and 12\ UTC. In Figure \ref{veri1} we {\myblack display} upper air verification for the LMCPF (dashed lines) and for the LETKF (solid lines). The different colors identify the different lead times, from one day up to one week. The first row shows the upper air temperature and the second row shows the u-component of wind. The first panel shows the Continuous Ranked Probability Score (CRPS), the second panel the Standard Deviation (SD), the third panel the Root Mean Square Error (RMSE) and the last panel shows the Mean (ME). For CRPS, SD and RMSE it is the aim to receive statistics as low as possible; for the Mean (=Bias) it is the goal to reach zero. {\myblack We used the same observations for verification in both experiments.}

Studying the results shown in Figure \ref{veri1}, we observe that the LMCPF shows slightly higher RMSE compared to the LETKF above 850 hPa. Forecast scores are nearly identical for LMCPF and LETKF for the upper air temperature, u-component of wind below 850hPa, i.e.,\ where the first guess statistics have been better for LMCPF. For upper air temperature at low levels and for high lead times the LMCPF errors are smaller than those of the LETKF. For the u-component of wind up to a height of approx.\ 100\ hPa the bias statistics of the LMCPF are better than the LETKF.

%-------------------------------------------------------------------------------
% 
%-------------------------------------------------------------------------------
In Figure \ref{veri2} we show the same statistics as in Figure \ref{veri1} focussing on relative humidity and upper air temperature for the comparison of LMCPF and LAPF. Here, it can be clearly seen that the LMCPF shows lower RMS errors than the LAPF for both variables and for all levels. For relative humidity the LMCPF is clearly better for the shorter lead times up to three days, but with less prominence it still outperforms the LAPF for the longer lead times up to one week. For the upper air temperature the RMSE statistics are clearly better for the LMCPF for all lead times. It is worth noting that the biases for the two particle filters show a quite similar behaviour. 

These results demonstrate that using particle uncertainty is an important ingredient for improving first guess and forecast scores of the particle filter. 

%-------------------------------------------------------------------------------
% 
%-------------------------------------------------------------------------------

%===============================================================================
%
%
%===============================================================================
\section{Conclusions}

In this work we develop the use of a Gaussian mixture within the framework of the Localized Adaptive Particle Filter (LAPF) introduced in \cite{Potthast19}, as an approximation to model and forecast particle uncertainty in the prior and posterior distributions. The filter, following earlier ideas of \cite{Hoteit08} and \cite{liu2016,liu2016b} constructs an analysis distribution based on localized Gaussian mixtures, whose posterior coefficients, covariances and means are calculated based on the prior mixture given by the ensemble first guess and the observations. The analysis step is completed by resampling and rejuvenation based on the LAPF techniques, leading to a Localized Mixture Coefficients Particle Filter (LMCPF). In contrast to the LAPF the LMCPF is characterized by a move or shift of the first guess ensemble towards the observations, which is consistent with the non-Gaussian posterior distribution based on a Bayesian analysis step, and where the size of the move is controlled by the size of the uncertainty of individual particles. 

We have implemented the LMCPF in the framework of the global ICON model for numerical weather prediction, operational at Deutscher Wetterdienst. Our reference system to test the feasibility of ideas and demonstrate the quality of the LMCPF is the LETKF implementation operational at DWD, which {\myred generates} initial conditions for the global ICON Ensemble Prediction System ICON-EPS. We have shown that the LMCPF runs stably for a month of global assimilations in operational setup and for a wide range of specific LMCPF parameters. Our investigation includes a study of the distribution of observations with respect to the ensemble mean and statistics of the distance of ensemble members to the projection of the observations into ensemble space. We also study the average size of particle moves when uncertainty is employed for individual Gaussian particles within the LMCPF and provide an analytic explanation of the histogram shapes with a comparison to the eigenvalue distribution of the metrical tensors on which the particle weights are based.

We show that the upper air {\em first guess errors} of the LMCPF and LETKF during the assimilation cycle are very similar within a range of plus-minus 1-3\%, with the LMCPF being better below 850 hPa and the LETKF being better above. {\em Forecast scores} for a time-period of one month have been calculated, demonstrating as well that the RMSE of the ensembles is comparable for upper air temperature and wind fields (2-3\%). The size of the mean spread of the LMCPF strongly depends on parameter choices and is usually stable after a spin-up period. 

In several shorter case studies we demonstrate that by varying the parameter choices, we can achieve better first guess RMSE for the LMCPF in comparison to the LETKF, which shows that for very short range forecasts the quality of the method can be comparable to or better than that of the LETKF. While reaching a break-even point for operational scores with a new method establishes an important mile-stone, we need to note that there are many open and intricate scientific questions here with respect to the choice of parameters for the Gaussian mixture and their inter-dependence as well as about the control of an optimal and correct ensemble spread both in the analysis cycle and for the forecasts.   

Overall, with the LMCPF we demonstrate significant progress compared to the localized adaptive particle filter (LAPF) for numerical weather prediction in an operational setup, demonstrating that the LMCPF has reached a stability and quality comparable to that of the LETKF, while allowing and taking care of diverse non-Gaussian distributions in its analysis steps. 

{\myblack Clearly, there is much more work to be done. The automatic choice of current tuning parameters is an important topic. Also, in further steps we will take a look at the quality control. Currently, the LMCPF and the LETKF are using the same observation quality control, but the LMCPF seems to need a more accurat approach. Furthermore, we have implemented the LAPF and LMCPF in the Lorenz 63 and Lorenz 96 models and are studying the characteristics of the particle filters in low-dimensional systems.}

%===============================================================================
%
%
%===============================================================================
\section*{Acknowledgements}

The research has been supported by the {\em Innovation in Applied Research and Development} (IAFE)
grant of the German Ministry for Transport and Digital Infrastructure BMVI. 
The authors thank Dr. Harald Anlauf and Dr. Andreas Rhodin from DWD for much support working on the 
DACE data assimilation coding environment.

\bibliographystyle{unsrtnat}
\bibliography{literature}  %%% Uncomment this line and comment out the ``thebibliography'' section below to use the external .bib file (using bibtex) .

\end{document}

%% file: mysty.tex
\usepackage{amsmath}
\usepackage{amsthm}
\usepackage{amsfonts}
\usepackage{amsxtra}
\makeindex

\newcommand{\cG}{{\mathcal G}}

\newcommand{\tx}{\tilde{x}}

\newcommand{\tL}{\tilde{L}}

\newcommand{\bd}{{\bf d}}
\newcommand{\bA}{{\bf A}}
\newcommand{\bB}{{\bf B}}

\newcommand{\bD}{{\bf D}}

\newcommand{\bG}{{\bf G}}
\newcommand{\bH}{{\bf H}}

\newcommand{\bI}{{\bf I}}

\newcommand{\bK}{{\bf K}}

\newcommand{\bN}{{\bf N}}

\newcommand{\bR}{{\bf R}}

\newcommand{\bW}{{\bf W}}
\newcommand{\bX}{{\bf X}}
\newcommand{\bY}{{\bf Y}}

\newcommand{\ox}{\overline{x}}
\newcommand{\oy}{\overline{y}}

\newcommand{\N}{\mathbb{N}}
\newcommand{\R}{\mathbb{R}}

\newcommand{\norm}[1]{|| #1||}

\newcommand{\ba}{\begin{array}}
\newcommand{\ea}{\end{array}}
\newcommand{\be}{\begin{equation}}
\newcommand{\ee}{\end{equation}}
\newcommand{\bea}{\begin{eqnarray}}
\newcommand{\eea}{\end{eqnarray}}
\newcommand{\beq}{\begin{equation}}
\newcommand{\eeq}{\end{equation}}
\newcommand{\bqt}{\begin{quote}}
\newcommand{\eqt}{\end{quote}}
\renewcommand{\span}[1]{{\rm span}\{ #1 \}}

\numberwithin{equation}{section}